\documentclass[a4paper,fleqn,usenatbib]{mnras}

\usepackage[T1]{fontenc}
\usepackage{ae,aecompl}
\usepackage{graphicx}	% Including figure files
\usepackage{amsmath}	% Advanced maths commands
\usepackage{amssymb}	% Extra maths symbols
\usepackage{color}

\newcommand{\msol}{\mbox{M$_{\odot}$}} 
 
\newcommand{\msolyr}{\mbox{M$_{\odot}$}\,yr$^{-1}$} 
\newcommand{\mdot}{$\dot{M}$}
\newcommand{\lsol}{\mbox{L$_{\odot}$}} 
 
\newcommand{\kks}{K km s$^{-1}$} 
\newcommand{\ks}{km s$^{-1}$}

\def\beginthetable{ \begin{table*} }
\def\endthetable{ \end{table*} }

\title[CO and mass loss of Bulge OH/IR stars]{On the detection of CO and mass
loss of Bulge OH/IR stars}
\author[Joris Blommaert et al.]{
J.A.D.L.~Blommaert,$^{1}$\thanks{E-mail: joris.blommaert@vub.be}
        M.A.T.~Groenewegen,$^{2}$
        K.~Justtanont,$^{3}$
        and L.~Decin,$^{4}$
\\
$^{1}$Astronomy and Astrophysics Research Group, Department of Physics and Astrophysics, \\ 
Vrije Universiteit Brussel, Pleinlaan 2, 1050 Brussels, Belgium \\
$^{2}$Koninklijke Sterrenwacht van Belgi\"e, Ringlaan 3, 1180 Brussel, Belgium \\
$^{3}$Chalmers University of Technology, Onsala Space Observatory, S-43992 Onsala, Sweden \\
$^{4}$Instituut voor Sterrenkunde, K.U. Leuven, Celestijnenlaan 200D, 3001 Leuven, Belgium
}

 \date{Accepted XXX. Received YYY; in original form ZZZ}

\pubyear{2018}

\begin{document}
\label{firstpage}
\pagerange{\pageref{firstpage}--\pageref{lastpage}}
\maketitle

\begin{abstract}
 
We report on the succesful search for CO~(2-1) and (3-2) emission associated  
with OH/IR stars in the Galactic Bulge. We observed a sample of eight
extremely red  AGB stars with the APEX telescope and detected seven. The 
sources were selected at sufficient high galactic latitude to avoid interference 
by interstellar CO,  which hampered previous studies of inner galaxy stars. To 
study the nature of our sample and the mass loss we constructed the SEDs from 
photometric data and Spitzer IRS spectroscopy. In a first step we apply 
radiative transfer  modelling to fit the SEDs and obtain luminosities and dust 
mass loss rates (MLR). Through  dynamical modelling we then retrieve the total 
MLR and the gas-to-dust ratios. We  derived variability periods of our stars.  
The luminosities range between approximately 4000 and 5500~L$_\odot$ and 
periods are below 700 days. The total MLR  ranges between $10^{-5}$ and 
$10^{-4}$ M$_{\odot}$  yr$^{-1}$. Comparison with evolutionary models shows  
that the progenitor mass $\approx 1.5$~M$_{\odot}$, similar to the Bulge 
Miras and are of intermediate age (3~Gyr). The gas-to-dust ratios are  between 
100 and 400 and are similar to what is found for OH/IR stars in the  galactic 
Disk. One star, IRAS~17347--2319, has a very short period of approximately 
300~days which may be decreasing further. It may belong to a class of Mira 
variables with  a sudden change in period as observed in some Galactic objects. 
It would be the first example of an OH/IR star in this class and deserves 
further follow-up observations.

\end{abstract}


% -------------------------------------------------------------------------------------------
% -------------------------------------------------------------------------------------------
\section{Introduction}
% -------------------------------------------------------------------------------------------
% -------------------------------------------------------------------------------------------
% -------------------------------------------------------------------------------------------
% -------------------------------------------------------------------------------------------

Low to intermediate mass stars (0.8 $\lesssim M_* \lesssim 8$~M$_{\odot}$) will 
ultimately end their lifes on the Asymptotic Giant Branch (AGB)
\citep{vassi93,habing1996}. 
Two striking characteristics of the AGB are the variability of the stars and
the mass loss. Different types of large amplitude variables are classified on
the 
basis of the amplitude: Semi-Regular Variables, Miras and OH/IR stars, where the
latter have the largest amplitudes (1 magnitude bolometric) and periods of several
hundred days. In the final phases on the AGB, the mass loss is the dominant 
process which will determine the AGB lifetime and its ultimate luminosity.  The 
mass-loss rates (MLRs) range approximately from $10^{-8}$  up to 
$10^{-4}$ M$_\odot$/yr.
Although the mass loss for these stars is already well known for many years 
%(for a review on AGB, see Habing 1996) 
there is still no firm understanding of what triggers the mass loss.
It is believed that, through large amplitude variability, the outer parts of the 
atmosphere are cool and the density is high enough to start dust formation. 
Radiation pressure on the grains drives these outwards, dragging with them the 
gas creating a slow ($\approx 15$ \ks) but strong stellar wind
\citep{Goldreich1976}. Through their mass loss, these stars provide a significant 
contribution to the gas and dust mass returned to the interstellar medium.  

When nearing the tip of the AGB, stars will start experiencing thermal pulses 
(a.k.a. helium shell flashes). 
The thermal pulses can lead
to the change of chemical type from the originally oxygen-rich star to either a 
S-type (C/O $\sim 1$) or carbon star (C/O $> 1$) through dredge-up of nuclear 
processed material to the surface \citep{Iben1975}. The change of chemical type 
is metallicity and stellar mass dependent. Stars with approximately solar 
metallicity and below two
or above four \msol\ are expected to remain oxygen-rich \citep{Marigo2013}. 

The OH/IR stars are the subset of AGB stars with the highest MLRs $> 10^{-5}$
M$_\odot$/yr observed \citep{Baud1983}. Such high MLR are significantly higher 
than the stellar 
mass loss description by \cite{Reimers1975} and are often called a {\it
superwind}, a term introduced by \cite{Renzini1981} to describe the MLR needed
to explain the characteristics of Planetary Nebulae.

The dust formed in the circumstellar shell completely obscures
the photospheric radiation and re-radiates it at infrared wavelengths
\citep{bedijn87}. The OH part 
of the name comes from the fact that in most cases of these infrared stars OH maser 
emisson, originating in a circumstellar thin shell, is detected. OH/IR stars are 
mostly found through either ``blind'' OH surveys that searched the galactic plane at the 18 cm 
radio line \citep[e.g.,][]{baud81,Sevenster1997} 
%, Lindqvist_obs1992, Blommaert1994, Sevenster1997, Sevenster2001}) 
or through a dedicated search on cool infrared sources with colours typical 
for a few 100 K temperature dust shell \citep{telintel1991}. A recent database
of circumstellar OH masers can be found in \citet{Engels2015}.

In the literature, OH/IR stars are often associated with more massive AGB 
stars. Well known studied examples of these are the OH maser sources near the 
galactic plane, 
like OH~26.5$+0.6$ (e.g. \cite{vanlangevelde90}). These stars have high 
luminosities well above 10,000~$\lsol$  and periods larger than a 1,000 days. 
Studies of OH/IR stars in the Bulge \citep{WilHarm1990,JEE2015} and the IRAS 
based study by \cite{Habing1988} of galactic disk OH/IR stars, find however 
luminosity distributions peaking at approximately 5,000~\lsol, expected to 
have relatively low mass progenitors below 2~\msol. 

Selecting Bulge stars provides the advantage of a relatively well known 
distance within our Galaxy. Generally the Bulge stellar population is considered 
to be
old \citep{Renzini1994, Zoccali2003, evelien2009}, however several studies also
indicate the presence of intermediate age stars \citep{vLoon2003, GroenBlom2005}. 
The question on the nature of the Bulge OH/IR stars 
is part of our analysis and will be discussed in Section~\ref{sec:population}.

The mass loss in AGB stars is studied by several means \citep{veenolofsson1990, 
Olofsson2003} of which infrared  studies of the circumstellar dust and the 
(sub-) millimetre detection of CO transitions are the most frequently used. 
In this study we combine the two techniques allowing to compare two independent
techniques and to study the gas-to-dust ratio, which is expected to be 
metallicity dependent. Earlier efforts to observe CO emission from AGB stars in 
the inner Galaxy had only limited success because of the interference
of interstellar CO emission along the galactic plane \citep{Winnberg2009, 
Sargent2013}, even though they used interferometric techniques. 
They selected OH/IR stars close to the galactic centre and plane respectively, 
which have different star formation histories than the Bulge
\citep{Launhardt2002, GenzelGC2010}. To avoid the galactic plane ISM 
interference we selected a population of OH/IR stars from the Bulge  
at higher latitudes. The sample selection of our paper is described in the 
following Section. We then continue with a description of the CO observations and 
data in Section~\ref{sec: data}. The results from the modelling of the IR and CO 
data are given in the
``Analysis'' Section~\ref{sec: analysis} \& \ref{sec:comparison}. The resulting characteristics are 
described in the Section~\ref{sec: characteristics}, followed by discussions on 
the Bulge
population of OH/IR stars and the superwind MLR in Section~\ref{sec:population}.

\section{Sample selection and description} 
\label{sec: sample}
% -------------------------------------------------------------------------------------------
% -------------------------------------------------------------------------------------------

The eight sources in this study are taken from a larger sample of fifty-three Galactic 
Bulge AGB stars which were selected to study the dust formation in the circumstellar
shell of oxygen-rich AGB stars \citep{Blommaert2007}.
The stars in the original sample were selected on the basis of infrared colours 
(observed with the ISO and/or IRAS satellites) to represent the whole range in MLR 
observed on the AGB, from naked stars with no observed mass loss up to OH/IR stars 
with MLRs in the order of $10^{-4}\,{\rm M}_{\odot} / {\rm yr}$.
Different studies of this sample were performed and presented in several papers: the 
dust content through Spitzer-IRS spectroscopy \citep{vanhollebeke2007, golriz14}, 
groundbased spectroscopy and photometry, including a monitoring programme to determine 
the variability \citep{vanhollebeke2007}(vH2007 from now on) 
and a high resolution near-infrared spectroscopic study of the abundances 
\citep{Uttenthaler2015}.  

From this sample we selected those with the reddest colours and thus also likely 
the stars with the 
highest MLRs ($\sim 10^{-4}\,{\rm M}_{\odot} / {\rm yr}$). The stars were
detected in the IRAS survey and originally studied in \citet{WilHarm1990}. 

We searched for counterparts of our sources in the OH maser database created by
\citet{Engels2015} which is considered complete for the published 1612 MHz maser
detections until the end of 2014.  
Seven out of our eight sources were searched for the OH (1612~MHz) maser 
emission and were detected by \citet{telintel1991, David1993, Sevenster1997}. 
The velocities of the OH maser 
emission peaks are given in Table~\ref{tab: sources}. For IRAS 17251--2821 two possible 
OH maser sources were detected and we give the observed velocities for both. 
The stars have absolute galactic 
latitudes above 2 degrees (except IRAS 17382-2830 with $b = 1.01^\circ$), which limits
the interference by interstellar CO and increases the chance to detect the 
circumstellar CO emission. Two sources are also detected in the infrared ISOGAL 
survey \citep{Omont2003}. 
As in the other papers on the Bulge sample, they are refered to with their ISOGAL name, 
these names have also been included in Table~\ref{tab: sources}.

\begin{table*}

\caption{Target list. } 

\label{tab: sources}

 \begin{tabular}{  l c c c c c c c}
  \hline
 & & & & & & & \\
  IRAS name   &  Right Ascension & Declination   & $l$ & $b$     & OH (peak velocities)   &  A$_V$ & ISOGAL name\\
               & (J2000)         & (J2000)       & (deg) & (deg) & LSR (\ks)             &  mag    &      \\
  \hline 
 & & & & & & & \\
17251-2821  & 17 28 18.60 & -28 24 00.4  & 358.41 & 3.49 & $-$181.0, $-$149.0$^a$ & 3.17 & \\
		&		&		&	&    &  $-$246.3, $-$227.5 & & \\
17276-2846  & 17 30 48.29 & -28 49 01.7 & 358.41 & 2.80 & $-$71.2, $-$40.4$^c$ &  4.27 & \\
17323-2424   & 17 35 25.92 & -24 26 30.5 & 2.61 & 4.31 & +70.0, +41.8$^c$ & 4.95 & \\
17347-2319  & 17 37 46.28 & -23 20 53.4  & 3.83 & 4.44 & +74.1, +90.1$^a$ &  4.29 & \\
17382-2830 & 17 41 22.59 & -28 31 48.0 & 359.86 & 1.01 & $-$68.7,   $-$33.7$^c$ & 5.19 & J174122.7-283146 \\
17413-3531  & 17 44 43.46 & -35 32 34.1 & 354.26 & $-$3.28 & --$^a$  & 2.44 & \\
17521-2938 & 17 55 21.80 & -29 39 12.9 & 0.47 & $-$2.19 & $-$88.5, $-$56.0$^b$ & 2.60 & J175521.7-293912 \\
18042-2905  & 18 07 24.39 & -29 04 48.0 & 2.27 & $-$4.19 &+39.2,  +69.1$^a$ & 1.48 & \\
\hline
\end{tabular}

Notes:  Positions taken from AllWISE \citep{cutri2013}. 
Velocities of the OH maser peak emission given with respect to the Local Standard of Rest (LSR). 
a: \cite{telintel1991}, b: \cite{David1993}, c: \cite{Sevenster1997}. 
Visual extinctions, A$_V$, are taken from \cite{vanhollebeke2007}, ISOGAL names are from \cite{Omont2003}.

\end{table*}

\begin{figure*}
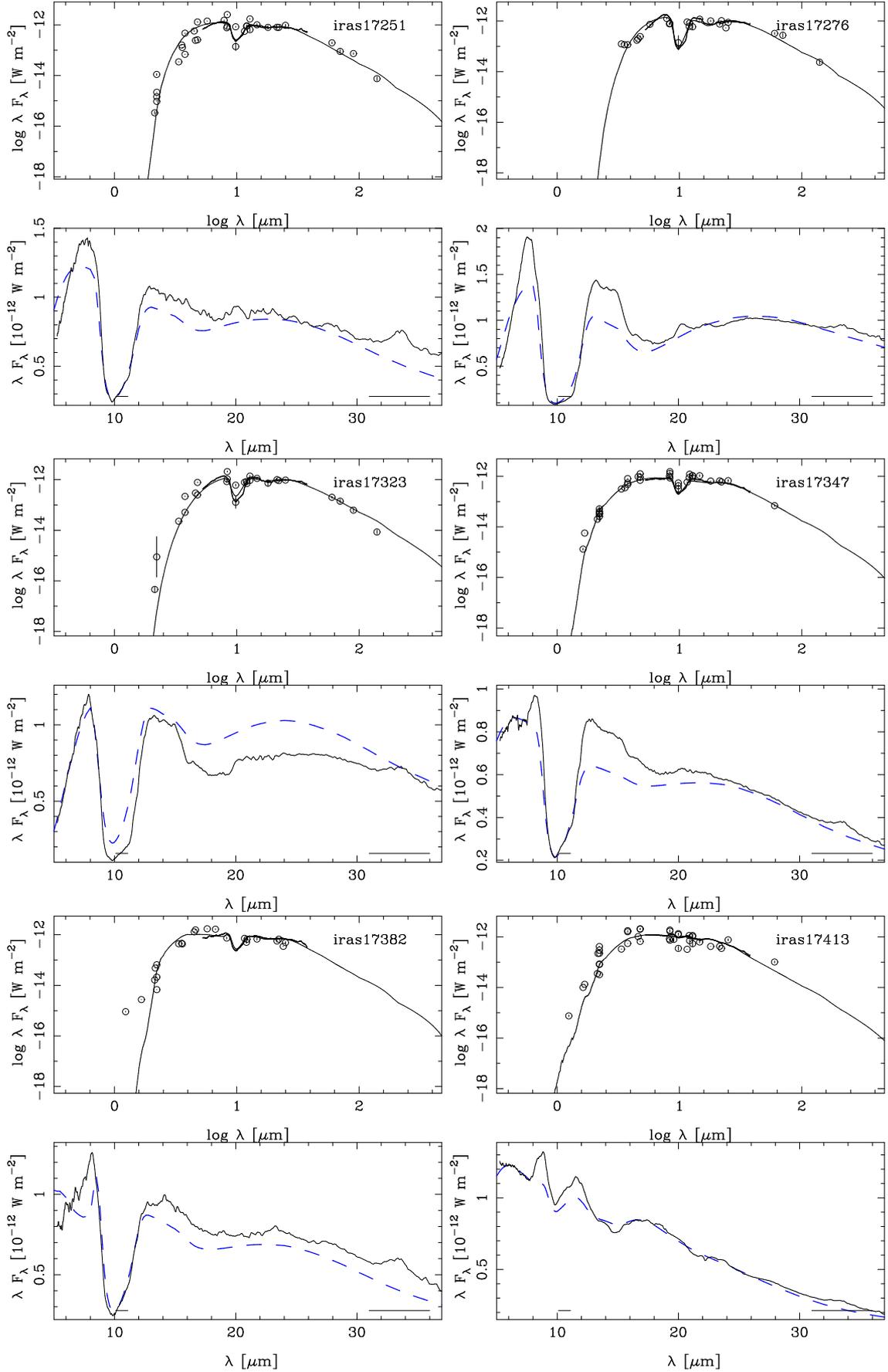

\label{fig: seds}

\begin{minipage}{0.42\textwidth}
\resizebox{\hsize}{!}{\includegraphics{iras17251_sed.ps}}
\end{minipage}
\begin{minipage}{0.42\textwidth}
\resizebox{\hsize}{!}{\includegraphics{iras17276_sed.ps}}
\end{minipage}

\begin{minipage}{0.42\textwidth}
\resizebox{\hsize}{!}{\includegraphics{iras17323_sed.ps}}
\end{minipage}
\begin{minipage}{0.42\textwidth}
\resizebox{\hsize}{!}{\includegraphics{iras17347_sed.ps}}
\end{minipage}

\begin{minipage}{0.42\textwidth}
\resizebox{\hsize}{!}{\includegraphics{iras17382_sed.ps}}
\end{minipage}
\begin{minipage}{0.42\textwidth}
\resizebox{\hsize}{!}{\includegraphics{iras17413_sed.ps}}
\end{minipage}

\caption[]{
Photometry and Spitzer-IRS spectra with the model fits (black full line vs 
photometry points and dashed blue vs IRS spectra), see
Sections~\ref{sec: irdata} \& \ref{sec:IRmodel}. The horizontal lines indicate
wavelength ranges with forsterite bands, which were discarded in
the modelling.
}
\end{figure*}
 
\renewcommand{\thefigure}{\arabic{figure} (Cont.)}
\addtocounter{figure}{-1}
\begin{figure*}
\begin{minipage}{0.42\textwidth}
\resizebox{\hsize}{!}{\includegraphics{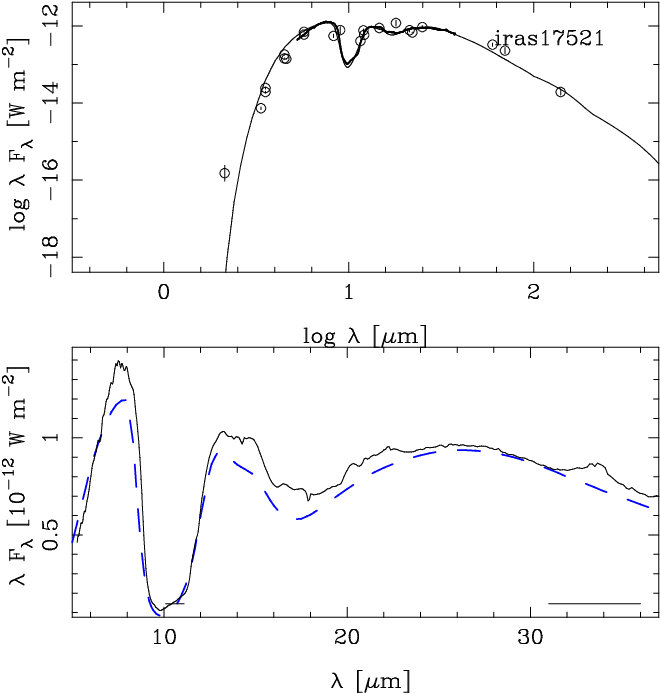}}
\end{minipage}
\begin{minipage}{0.42\textwidth}
\resizebox{\hsize}{!}{\includegraphics{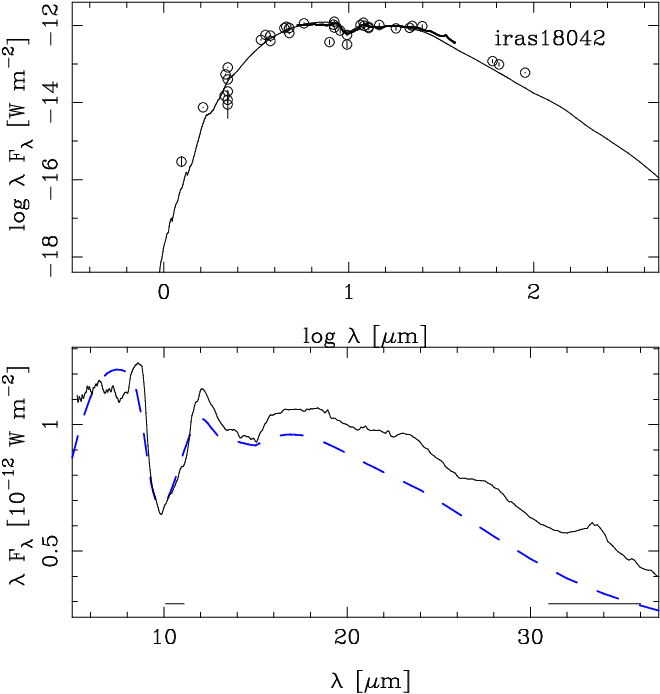}}
\end{minipage}
\caption[]{
}
\label{Fig-SEDcont}
\end{figure*}
\renewcommand{\thefigure}{\arabic{figure}}
	    
\cite{JEE2015} (JEE15 from now onwards) selected a sample of thirty-seven
Bulge IRAS 
sources with OH/IR star-like IRAS colours and 
modelled the spectral energy distributions (SED). We will compare our analysis 
with theirs in Section~\ref{sec:comparison}. A difference in the selection of
our sample, based on \citet{Blommaert2007}, with JEE15 is that all our sources
have IRAS 12~$\mu$m flux densities below 10~Jy. This limit was imposed as
\citet{WilHarm1990} considered these to be most likely Bulge members and not of the
galactic disk. JEE15 did not impose such a flux criterion.

\section{Data and observations description}
\label{sec: data}

\subsection{Spectral Energy Distribution data}
\label{sec: irdata}
We made use of VizieR \citep{VizieR} and data in the literature to collect photometric data at the 
position of the OH/IR stars to create their SEDs.
Two important aspects about our sources need to
be considered. One is the fact that our stars are highly variable and that the
data obtained from different catalogues and publications have not been observed 
at the same single epoch. Second, the Bulge is a high source-density area, so that 
confusion with nearby sources is a risk. The strategy followed was to start 
searching in the VizieR database 
for the nearest AllWISE counterpart \citep{cutri2013} of the IRAS position. The
WISE and IRAS surveys overlap in the mid-infrared wavelength regime, where 
fewer sources are detected, limiting the chance of a wrong association. 
The WISE sources selected in this manner have a [W3] $-$ [W4] colour of about 2
magnitudes, consistent with the IRAS colours.
The AllWISE positions are accurate up to 50~mas. In a second step, a search 
area of $ 4^{\prime\prime}$ radius around the position was used to search 
in other infrared catalogues. In case of finding more than one source, the 
reddest object is selected. The data is given in Table~\ref{tab: photom}.

\begin{table*}

\centering

 \caption{ Photometric data of our targets.}
 
\scriptsize
\begin{tabular}{ c c c c c c c c c}
  \hline
 & & & & & & & & \\
% & J & H & K & nbL & K & L & M & N1 & N2 & N3 & irac1 & irac2 & irac3 & W1 & W2
% & W3 & W4 \\
  Filter   &  17251--2821 & 17276--2846 & 17323--2424 & 17347--2319 & 17382--2830 & 17413--3531 & 17521--2938 & 18042--2905 \\
%              & 1.2 & 1.65 & 2.2 & 2.2 & "3.6" &  &  &  &  & 3.35  & 4.6 & 11.6
%          & 22.1 \\
 & & & & & & & & \\
  \hline 
% & & & & & & & & \\
\underline{CASPIR (mag)}  & & & & & & & & \\
%   & & & & & & & & \\
%J  &                  & & & & & & &  \\
H  &                  & & & 13.78 $\pm$ 0.01 &            & 12.85 $\pm$ 0.01 & &  \\
K  & 14.02 $\pm$ 0.01 & & & 10.86 $\pm$ 0.01 & 10.33 $\pm$ 0.01 &  8.99 $\pm$ 0.01 & & 10.09 $\pm$ 0.01 \\ 
K  & 14.47 $\pm$ 0.01 & & & 11.09 $\pm$ 0.01 & 11.56 $\pm$ 0.01 & 10.09 $\pm$ 0.01 & & 10.87 $\pm$ 0.01 \\ 
K  & 14.92 $\pm$ 0.01 & & & 11.21 $\pm$ 0.01 & 12.79 $\pm$ 0.01 & 11.10 $\pm$ 0.01 & & 11.65 $\pm$ 0.01 \\ 
nbL & 8.11 $\pm$ 0.02 & & &  7.02 $\pm$ 0.02 &  6.80 $\pm$ 0.02 &            & &         \\
\underline{ESO (mag)}  & & & & & & & & \\
% & & & & & & & & \\
K & 12.27 $\pm$ 0.08 &           & 15.0  $\pm$ 0.2  & 10.6  $\pm$ 0.1  & &  8.69 $\pm$ 0.07 & & 12.5 $\pm$ 0.9  \\
K &                  &           &            & 10.98 $\pm$ 0.03 & & 10.06 $\pm$ 0.02 & & 12.2 $\pm$ 0.2  \\
K &                  &           &            & 11.1  $\pm$ 0.7  & &  8.33 $\pm$ 0.03 & &  \\
L &  6.53 $\pm$ 0.04 & 8.05 $\pm$ 0.02 &  7.36 $\pm$ 0.03 &  5.94 $\pm$ 0.02 & & 5.21 $\pm$ 0.04 & & 6.38 $\pm$ 0.02  \\
L &  8.62 $\pm$ 0.03 &           &  8.94 $\pm$ 0.05 &  6.36 $\pm$ 0.1  & & 6.37 $\pm$ 0.01 & & 6.73 $\pm$ 0.03  \\
L &                  &           &            &  6.53 $\pm$ 0.03 & & 5.10 $\pm$ 0.02 & & \\
M &  4.77 $\pm$ 0.05 & 6.59 $\pm$ 0.08 &  5.33 $\pm$ 0.04 &  4.81 $\pm$ 0.04 & & 4.30 $\pm$ 0.04 & & 5.23 $\pm$ 0.07 \\
M &  6.53 $\pm$ 0.07 &           &  6.56 $\pm$ 0.08 &  5.15 $\pm$ 0.03 & & 5.51 $\pm$ 0.04 & & 5.56 $\pm$ 0.05 \\
M &                  &           &            &  5.46 $\pm$ 0.05 & & 4.26 $\pm$ 0.06 & & \\
N1 & 2.24 $\pm$ 0.03 & 3.5 $\pm$ 0.1 & 2.46 $\pm$ 0.05 & 2.80 $\pm$ 0.03 & 2.59 $\pm$ 0.03 & & & 3.02 $\pm$ 0.03 \\
N1 & 3.4  $\pm$ 0.1  & 4.9 $\pm$ 0.7 & 3.19 $\pm$ 0.08 & 2.99 $\pm$ 0.07 & 3.6  $\pm$ 0.01 & & & 3.4  $\pm$ 0.1 \\
N1 &                 &         &           & 3.43 $\pm$ 0.09 & 2.70 $\pm$ 0.07 & & & \\
N2 & 2.95 $\pm$ 0.05 & 2.4 $\pm$ 0.3 & 3.3 $\pm$ 0.1   & 3.42 $\pm$ 0.03 & 2.44 $\pm$ 0.06 & & & 3.39 $\pm$ 0.05  \\
N2 & 4.9  $\pm$ 0.4  &         & 5.0 $\pm$ 0.6   & 3.7  $\pm$ 0.1  & 3.9  $\pm$ 0.2  & & & 4.0  $\pm$ 0.2 \\
N2 &                 &         &           & 4.0  $\pm$ 0.4  & 2.54 $\pm$ 0.07 & & &\\
N3 & 1.24 $\pm$ 0.05 &         & 1.48 $\pm$ 0.09 & 1.77 $\pm$ 0.04 & 1.7 $\pm$ 0.1 & & & 1.93 $\pm$ 0.07 \\
N3 & 2.3  $\pm$ 0.3  &         & 1.9  $\pm$ 0.2  & 1.8  $\pm$ 0.1  & 2.5 $\pm$ 0.3 & & & 2.0  $\pm$ 0.1 \\
N3 &                 &         &           & 2.0  $\pm$ 0.2  & 1.8 $\pm$ 0.1 & & &  \\
\underline{2MASS (mag)}  & & & & & & & & \\
%  & & & & & & & & \\
J          &  & &  &  & 16.57 $\pm$ 0.04 & & &  \\
H          &  & &  &  & 14.59 $\pm$ 0.02 & & &  \\
K$_{\rm s}$ &  & &  &  & 10.72 $\pm$ 0.01 & 9.06 $\pm$ 0.02 & & 10.59 $\pm$ 0.02 \\
\underline{VISTA (mag)}  & & & & & & & & \\
%          & & & & & & & & \\
J         &                  &     &                 &                 &                  & 16.75 $\pm$ 0.09 &                & 17.77 $\pm$ 0.30 \\
H         &                  &     &                 & 15.4  $\pm$ 0.1 &                  & 13.20 $\pm$ 0.01 &                & 13.51 $\pm$ 0.01 \\
K$_{\rm s}$ & 16.13 $\pm$ 0.15 &     & 18.3 $\pm$ 0.3 & 11.70 $\pm$ 0.01 & 11.92 $\pm$ 0.01 & 11.08 $\pm$ 0.01 & 17.0 $\pm$ 0.5 & 12.01 $\pm$ 0.01 \\
% & & & & & & & & \\
\underline{GLIMPSE (mag)}  & & & & & & & & \\
% & & & & & & & & \\
irac36 & 7.91 $\pm$ 0.03 & 8.27 $\pm$ 0.03 & & & 6.77 $\pm$ 0.06 &  &  9.96 $\pm$ 0.05 & 6.55 $\pm$ 0.05 \\
irac36 &                 &           & & &                 &  & 10.22 $\pm$ 0.06       & \\
irac45 & 5.80 $\pm$ 0.05 & 7.10 $\pm$ 0.03 & & & 4.91 $\pm$ 0.05 &  & 7.06  $\pm$ 0.04 & 5.37 $\pm$ 0.08 \\
irac45 &                 &           & & &                 &  & 7.31 $\pm$ 0.04       & \\
irac58 & 4.10 $\pm$ 0.02 & 4.81 $\pm$ 0.02 & & & 3.89 $\pm$ 0.02 &  & 4.87  $\pm$ 0.02 & 4.35 $\pm$ 0.04 \\
irac58 &                 &           & & &                 &  & 5.05 $\pm$ 0.03       & \\
irac80 & 3.00 $\pm$ 0.02 & 3.20 $\pm$ 0.06 & & &                 &  &                  & 4.55 $\pm$ 0.15 \\
% & & & & & & & & \\
\underline{AllWISE (mag)}  & & & & & & & & \\
% & & & & & & & & \\
WISE1 (3.4~$\mu$m) & 9.73 $\pm$ 0.03 & 8.35 $\pm$ 0.02 & 10.20 $\pm$ 0.03 & 7.35 $\pm$ 0.03 & 6.98 $\pm$ 0.03 & 7.29 $\pm$ 0.03 & 11.43 $\pm$ 0.08 & 7.02 $\pm$ 0.03 \\
WISE2 (4.6~$\mu$m) & 6.63 $\pm$ 0.02 & 6.86 $\pm$ 0.02 &  6.42 $\pm$ 0.02 & 5.15 $\pm$ 0.06 & 4.62 $\pm$ 0.04 & 5.07 $\pm$ 0.04 &  7.24 $\pm$ 0.02 & 5.19 $\pm$ 0.08 \\
WISE3 (12~$\mu$m) & 2.75 $\pm$ 0.01 & 2.21 $\pm$ 0.01 &  2.38 $\pm$ 0.01 & 2.70 $\pm$ 0.01 & 2.48 $\pm$ 0.02 & 3.35 $\pm$ 0.01 &  3.10 $\pm$ 0.01 & 2.09 $\pm$ 0.01 \\
WISE4 (22~$\mu$m) & 0.34 $\pm$ 0.02 & 0.06 $\pm$ 0.01 &  0.18 $\pm$ 0.02 & 0.65 $\pm$ 0.01 & 0.53 $\pm$ 0.02 & 1.23 $\pm$ 0.02 &  0.52 $\pm$ 0.02 & 0.12 $\pm$ 0.01 \\
% & & & & & & & & \\
\underline{ISOGAL (mag)}   & & & & & & & & \\
% & & & & & & & & \\
LW2 (7~$\mu$m)  & & & & & 3.47 $\pm$ 0.01 & & & \\
LW3 (15~$\mu$m) & & & & & 1.54 $\pm$ 0.03 & & & \\
% & & & & & & & & \\
\underline{IRAS (Jy)}   & & & & & & & & \\
% & & & & & & & & \\
F12 & 3.6 $\pm$ 0.4 & 2.5 $\pm$ 0.3 & 3.4 $\pm$ 0.3 & 3.6 $\pm$ 0.4 & 2.0 $\pm$ 0.2 & 5.0 $\pm$ 0.5 & 3.1 $\pm$ 0.3 & 4.7 $\pm$ 0.5 \\
F25 & 8.5 $\pm$ 0.9 & 7.9 $\pm$ 0.8 & 8.5 $\pm$ 0.9 & 5.9 $\pm$ 0.6 & 4.3 $\pm$ 0.4 & 6.8 $\pm$ 0.7 & 8.2 $\pm$ 0.8 & 8.4 $\pm$ 0.8 \\
F60 & 4.3 $\pm$ 0.4 & 7.1 $\pm$ 0.7 & 4.4 $\pm$ 0.4 & 1.5 $\pm$ 0.2 &               & 2.2 $\pm$ 0.2 & 7.1 $\pm$ 0.7 & 2.6 $\pm$ 0.3 \\
% & & & & & & & & \\
\underline{MSX (Jy)}  & & & & & & & & \\
% & & & & & & & & \\
A (8.28)  & 2.1 $\pm$ 0.2 & 2.1 $\pm$ 0.2 & 2.3 $\pm$ 0.2 & 2.8 $\pm$ 0.3 & 2.1 $\pm$ 0.2 & 2.3 $\pm$ 0.2 & 1.5 $\pm$ 0.5 & 2.9 $\pm$ 0.3 \\
C (12.13) & 3.7 $\pm$ 0.4 & 3.4 $\pm$ 0.3 & 2.9 $\pm$ 0.3 & 4.9 $\pm$ 0.5 & 2.7 $\pm$ 0.3 & 2.9 $\pm$ 0.3 & 2.4 $\pm$ 0.2 & 4.0 $\pm$ 0.4 \\
D (14.65) & 4.9 $\pm$ 0.5 & 5.8 $\pm$ 0.6 & 5.5 $\pm$ 0.6 & 5.2 $\pm$ 0.5 & 3.3 $\pm$ 0.3 & 3.1 $\pm$ 0.3 & 4.3 $\pm$ 0.4 & 5.0 $\pm$ 0.5 \\
E (21.34) & 5.8 $\pm$ 0.6 & 5.4 $\pm$ 0.5 & 7.1 $\pm$ 0.7 & 4.7 $\pm$ 0.5 & 4.0 $\pm$ 0.4 & 3.0 $\pm$ 0.5 & 5.5 $\pm$ 0.6 & 6.2 $\pm$ 0.6 \\
% & & & & & & & & \\
\underline{MIPS (mag)}  & & & & & & & & \\
% & & & & & & & & \\
mips24 & & 0.58 $\pm$ 0.02 & & & 1.05 $\pm$ 0.02 & & & \\
% & & & & & & & & \\
\underline{AKARI (Jy)}   & & & & & & & & \\
% & & & & & & & & \\
S9  &               &         &               &                 &  & 2.7  $\pm$ 0.6  & 2.59 $\pm$ 0.60 & 2.47 $\pm$ 0.40 \\
S18 & 4.7 $\pm$ 0.9 & 5.9 $\pm$ 0.4 & 4.3 $\pm$ 0.2 & 3.79 $\pm$ 0.01 &  & 2.47 $\pm$ 0.02 & 6.94 $\pm$ 1.27 & 4.89 $\pm$ 0.62 \\
S65 &               &               &               &                 &  &                 &                 & 2.09 $\pm$ 0.22 \\
S90 & 2.2 $\pm$ 0.2 &               & 1.9 $\pm$ 0.4 &                 &  &                 &                 & 1.78 $\pm$ 0.08 \\
% & & & & & & & & \\
\underline{PACS (Jy)}   & & & & & & & & \\
% & & & & & & & & \\
 70 &  2.1 $\pm$ 0.3 & 6.4 $\pm$ 1.3 & 3.3 $\pm$ 0.7 &  & & & 5.3 $\pm$ 1.1 & \\
140 &  0.4 $\pm$ 0.1 & 1.1 $\pm$ 0.2 & 0.4 $\pm$ 0.1 &  & & & 0.9 $\pm$ 0.2 & \\
\hline
\end{tabular}

\label{tab: photom}

\end{table*}

We have complemented the VizieR data with the $J, H, K$, nb$L$ averaged photometry 
from the vH2007 monitoring programme  
obtained at the Mount Stromlo observatory. We also included the ESO photometric 
2 -- 13~$\mu$m data obtained by \citet{WilHarm1990}. 
Finally, public DR4 data from the VVV Survey \citep{Minniti2010} was included\footnote{see \url{http://horus.roe.ac.uk/vsa/index.html}}.
IRAS17521 and IRAS17323 are not listed in the source catalog, but are visible on the $K$-band image 
and the magnitudes have been estimated by scaling the flux (minus background) in a 3$\times$3 pixel region 
to that of a nearby catalogue star.

Finally half of our sample were also observed with the Herschel PACS spectrometer
\citep{pilbratt10,poglitsch10} in the open time programme ``Study of the cool 
forsterite
dust around evolved stars'' (OT2\_jblommae\_2). The flux densities given in 
Table~\ref{tab: photom} are the continuum levels at 70 and 140~$\mu$m of the 
central spaxel as obtained from archive pipeline product v14. The flux densities 
were corrected for the missing part of the point spread function (PSF). The formal
uncertainties of the PACS spectrometer flux calibration is 15\%. 

Figure~1
%\ref{fig: seds} krijg hier steeds 2
shows the obtained SEDs and the model  fits obtained (see Section
\ref{sec:IRmodel}). The figure also includes the Spitzer-IRS spectrum covering 
the 5-37~$\mu$m wavelength range and which are taken from \cite{golriz14}. All sources 
show typical SEDs for OH/IR stars, i.e. an optically thick silicate-rich dust 
shell with strong absorption features at 9.7 and 18~$\mu$m.

\subsection{CO Observation and data reduction} 
\label{sec: COdata}
The CO J= 2-1 and J= 3-2 transitions were observed with the APEX telescope located 
in the Atacama dessert in Chile \citep{APEX}.
The observations were obtained in service mode on 
September 11, 12, 13, November 10, 11, 12, 2011 (I17276, I17323, I17521, I18042) 
and
September 26, 27, 29, 30, 2012 (I17251, I17347, I17413, I17382).
Weather conditions varied but most observations were taken with a precipitable water 
vapour (PWV) between 0.7 and 1.3 mm for the J= 3-2, and between 1 and 2 mm for 
the J= 2-1 transition.
The APEX-1 and APEX-2 receivers of the Swedish Heterodyne Facility Instrument 
(SHeFI)\footnote{http://gard04.rss.chalmers.se/APEX\_Web/SHeFI.html}
\citep{Belitsky2006, Vassilev2008} were tuned to the CO J= 2-1 and 3-2 line, respectively. 
The beam size and the main-beam efficiency at these frequencies are 26.4\arcsec\ 
(FWHM), $\eta_{\rm mb}$= 0.73,
respectively, 17.3\arcsec, 0.75.
The XFFTS (eXtended bandwidth Fast Fourier Transform Spectrometer) backend 
(see \cite{Klein2012}) was connected to the receivers.
Wobbler switching was used with a throw of 50\arcsec. Regular observations of bright 
sources were performed to check the pointing and calibration. 

The data were reduced in CLASS\footnote{http://www.iram.fr/IRAMFR/GILDAS/}. 
Linear baselines were subtracted avoiding regions that were affected by 
interstellar contamination and the location
of the CO detection (or using the velocity range suggested by the OH maser 
emission line in case of a CO non-detection). Typical total integration times per
source were 40-50 minutes for the J=2-1 and 100-130 minutes for the J=3-2
transitions, leading to a RMS of $\approx 10$ mK for both transitions at a velocity
resolution of 1 \ks.

\begin{figure*}
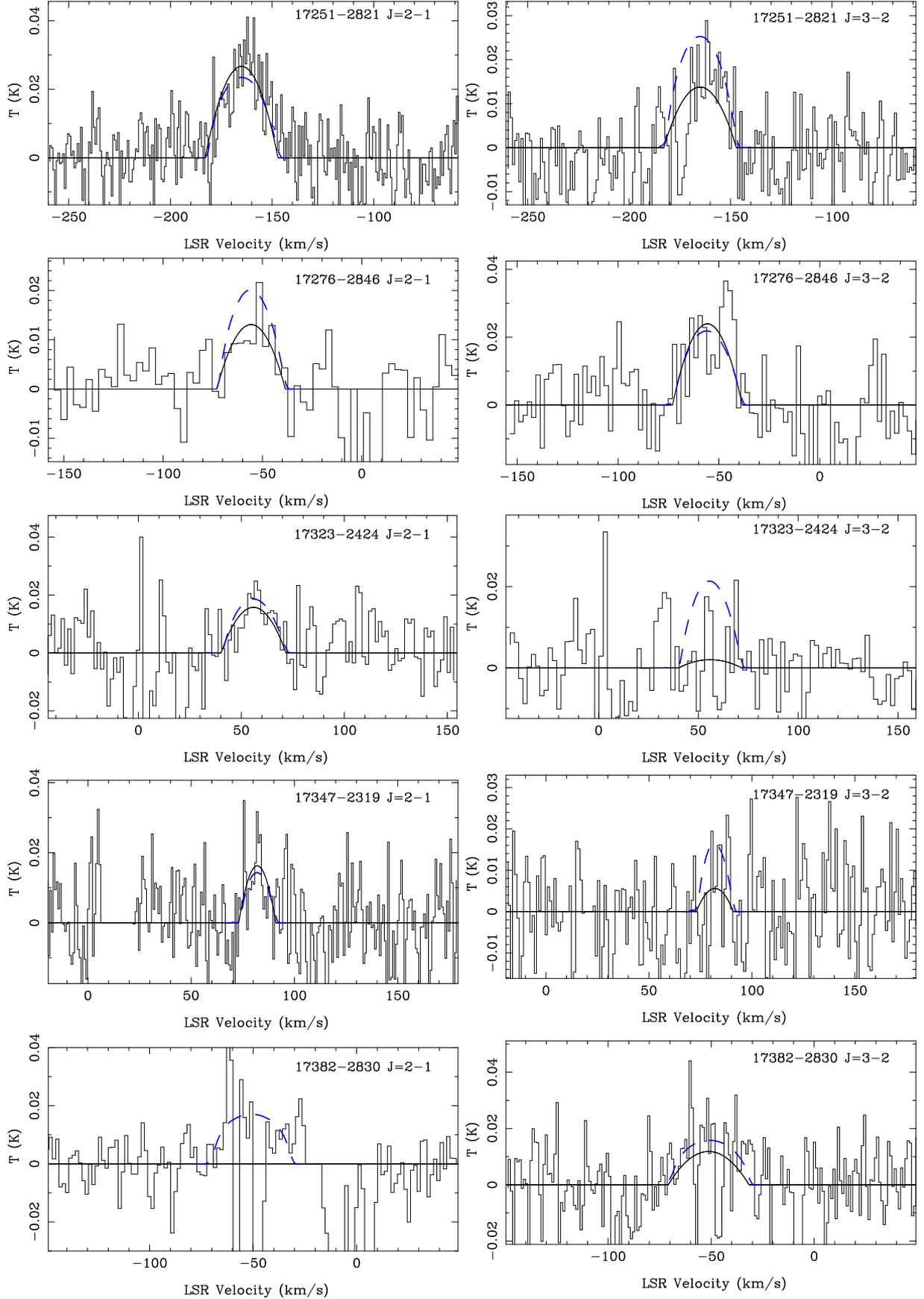


\begin{minipage}{0.45\textwidth}
\resizebox{\hsize}{!}{\includegraphics{profilefit_I17251_CO21_F.ps}}
\end{minipage}
\begin{minipage}{0.45\textwidth}
\resizebox{\hsize}{!}{\includegraphics{profilefit_I17251_CO32_F.ps}}
\end{minipage}

\begin{minipage}{0.45\textwidth}
\resizebox{\hsize}{!}{\includegraphics{profilefit_I17276_CO21_F.ps}}
\end{minipage}
\begin{minipage}{0.45\textwidth}
\resizebox{\hsize}{!}{\includegraphics{profilefit_I17276_CO32_F.ps}}
\end{minipage}

\begin{minipage}{0.45\textwidth}
\resizebox{\hsize}{!}{\includegraphics{profilefit_I17323_CO21_F.ps}}
\end{minipage}
\begin{minipage}{0.45\textwidth}
\resizebox{\hsize}{!}{\includegraphics{profilefit_I17323_CO32_F.ps}}
\end{minipage}

\begin{minipage}{0.45\textwidth}
\resizebox{\hsize}{!}{\includegraphics{profilefit_I17347_CO21_F.ps}}
\end{minipage}
\begin{minipage}{0.45\textwidth}
\resizebox{\hsize}{!}{\includegraphics{profilefit_I17347_CO32_F.ps}}
\end{minipage}

\begin{minipage}{0.45\textwidth}
\resizebox{\hsize}{!}{\includegraphics{profilefit_I17382_CO21_F.ps}}
\end{minipage}
\begin{minipage}{0.45\textwidth}
\resizebox{\hsize}{!}{\includegraphics{profilefit_I17382_CO32_F.ps}}
\end{minipage}

\caption[]{
The APEX CO (2-1) and (3-2) line spectra, together with the line fits (in black)
and the model predictions (dashed blue). For a description, see Sections~\ref{sec: irdata} \& 
\ref{sec:COmodel}.
}
\label{Fig-CO}
\end{figure*}

\renewcommand{\thefigure}{\arabic{figure} (Cont.)}
\addtocounter{figure}{-1}

\begin{figure*}
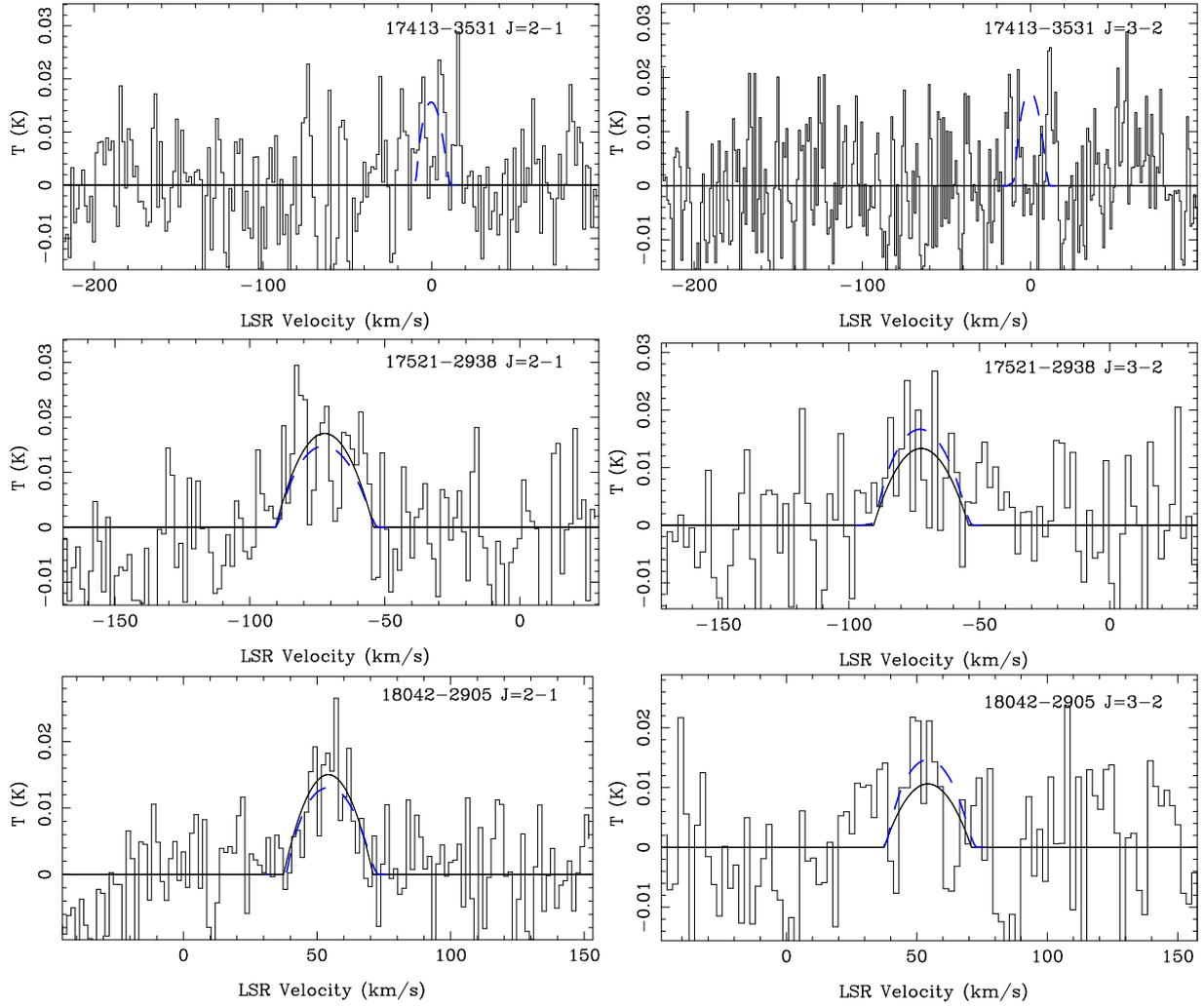


\begin{minipage}{0.45\textwidth}
\resizebox{\hsize}{!}{\includegraphics{profilefit_I17413_CO21_F.ps}}
\end{minipage}
\begin{minipage}{0.45\textwidth}
\resizebox{\hsize}{!}{\includegraphics{profilefit_I17413_CO32_F.ps}}
\end{minipage}

\begin{minipage}{0.45\textwidth}
\resizebox{\hsize}{!}{\includegraphics{profilefit_I17521_CO21_F.ps}}
\end{minipage}
\begin{minipage}{0.45\textwidth}
\resizebox{\hsize}{!}{\includegraphics{profilefit_I17521_CO32_F.ps}}
\end{minipage}

\begin{minipage}{0.45\textwidth}
\resizebox{\hsize}{!}{\includegraphics{profilefit_I18042_CO21_F.ps}}
\end{minipage}
\begin{minipage}{0.45\textwidth}
\resizebox{\hsize}{!}{\includegraphics{profilefit_I18042_CO32_F.ps}}
\end{minipage}

\caption[]{
No detection for IRAS~17413--3531, the blue line indicates
the predicted line strength from our dynamical modelling at a LSR velocity of 0
km $s^{-1}$. 
}
\label{Fig-COcont}
\end{figure*}
\renewcommand{\thefigure}{\arabic{figure}}

The resulting profiles are shown in Figure~\ref{Fig-CO} plotting main-beam temperatures 
against velocity (using the LSR as the velocity reference).

The profiles were fitted with our own Fortan version of the "Shell" profile 
available within CLASS/GILDAS software 
package\footnote{http://www.iram.fr/IRAMFR/GILDAS/doc/html/class-html/node38.html},
\begin{equation}
   P(V) = \frac{I}{\Delta V \; (1 + H /3)} \; \left(1 + 4 H \; \left(\frac{V - V_0}{\Delta V}\right)^2\right),
\end{equation}
where $V_0$ is the stellar velocity (in \ks), $I$ is the integrated intensity (in 
\kks), $\Delta V$ the full-width at zero intensity (in \ks, and the expansion 
velocity v$_{\rm exp}$ is taken as half that value), and $H$ the horn-to-center 
parameter. This parameter described the shape of the profile, with
$-1$ for a parabolic profile, $0$ for a flat-topped one, and $>0$ for a 
double-peaked profile.
In the fitting below, a parabolic profile was assumed for all cases. From the 
fitting we obtain the stellar radial and expansion velocities. However as the
CO profiles are relatively weak, we have chosen to use the OH velocity information
and keep the stellar and expansion velocities fixed. \cite{groen98} compared 
expansion velocities derived from OH and CO observations and found that generally
the CO 	profiles are 12\% wider and we applied this correction for our fitting.

The results are listed in Table~\ref{tab:CO_results}.
The errors in the parameters were estimated by a Monte Carlo simulation where the
intensity in every channel was varied according to a Gaussian with the observed 
rms noise (assuming the channels are independent), and the profile 
refitted. For sources where we could obtain an independent 
fitting of the CO profile, without using the OH
velocities we give the obtained velocities in the same Table. The "CO" and "modified OH"
velocities generally agree within the errors.

\beginthetable
%\begin{table}
\setlength{\tabcolsep}{0.7mm}
%\scriptsize
\footnotesize
\centering
\caption{CO data fitting results. }
\begin{tabular}{lcccccccccc}
\hline 
%  &  & & &  &  &  \\
IRAS  & $V_{\rm LSR}$  & $\Delta V$ & $T_{2-1,{\rm peak}} $ & $I_{2-1}$ &
            $T_{3-2,{\rm peak}} $ & $I_{3-2}$ & $V_{\rm LSR}(2-1)$  & $\Delta V (2-1)$ & $V_{\rm LSR}(3-2)$  & $\Delta V (3-2)$  \\
name      & (\ks)  & (\ks) & (K) & (K~\ks) & (K)  & (K~\ks) & (\ks)  & (\ks) &  (\ks)  & (\ks) \\
\hline
  &  & & &  &  & & &  &  & \\
17251-2821  & -165.0 & 35.8 & 0.027 $\pm$ 0.002 & 0.637 $\pm$ 0.042 & 0.014 $\pm$ 0.002 & 0.327 $\pm$ 0.047 & -162.3 $\pm$ 0.9 & 37.1 $\pm$ 2.4 & -160.0 $\pm$ 0.5 & 28.3 $\pm$ 3.3 \\
17276-2846  &  -55.8 & 34.5 & 0.013 $\pm$ 0.004 & 0.305 $\pm$ 0.090 & 0.024 $\pm$ 0.004 & 0.55  $\pm$ 0.10  &  -55.7 $\pm$ 2.0 & 31.7 $\pm$ 6.1 &  -53.6 $\pm$ 2.0 & 36.7 $\pm$ 5.9 \\
17323-2424  &   55.9 & 31.6 & 0.016 $\pm$ 0.005 & 0.336 $\pm$ 0.096 & -                 & -                 &   58.9 $\pm$ 1.5 & 28.0 $\pm$ 4.0 &         & \\
17347-2319  &   82.0 & 17.9 & 0.017 $\pm$ 0.003 & 0.198 $\pm$ 0.039 & 0.005 $\pm$ 0.004 & 0.065 $\pm$ 0.046 &       &  &   & \\
17382-2830  &  -51.2 & 39.2 & -                 & -                 & 0.012 $\pm$ 0.003 & 0.312 $\pm$ 0.086 &       &  &    & \\
17413-3531  &  - & - & - & - & - & - &   &  &    & \\
17521-2938  &  -72.3 & 36.4 & 0.017 $\pm$ 0.003 & 0.416 $\pm$ 0.067 & 0.014 $\pm$ 0.003 & 0.323 $\pm$ 0.075 &  -73.3 $\pm$ 2.2 & 47.3 $\pm$ 6.7 &  -75.0 $\pm$ 3.0 & 40.9 $\pm$ 8.0 \\
18042-2905  &   54.2 & 33.5 & 0.015 $\pm$ 0.002 & 0.338 $\pm$ 0.043 & 0.011 $\pm$ 0.003 & 0.236 $\pm$ 0.076 &   55.3 $\pm$ 1.3 & 28.9 $\pm$ 2.9 &   51.6 $\pm$ 3.9 & 28.5 $\pm$ 7.0 \\
\hline
\end{tabular}
\label{tab:CO_results}
%\end{table}
\\
Notes: 
The $V_{\rm LSR}$ and $\Delta V$ are taken from the OH observations. For the
sources where an independent fit of the CO profile was possible 
(see Section~\ref{sec: COdata}), the velocities are given
in the last 4 columns. 

\endthetable

\section{Analysis}
\label{sec: analysis}

\subsection{Modelling of the IR data}
\label{sec:IRmodel}

The models are based on the "More of DUSTY" (MoD) code \citep{groenewegen12} which uses a
slightly updated and modified version of the {\em DUSTY} dust radiative transfer (RT) code \citep{Dusty1999}
as a subroutine within a minimization code.
The code determines the best-fitting dust optical depth, luminosity, dust temperature
at the inner radius and the slope, $p$, of the density distribution ($\rho \sim r^{-p}$)
by fitting photometric data and spectra (and visibility data and 1D intensity profiles,
but these data are not available for the sample considered here).
The code minimizes a $\chi^2$ based on every available photometric and spectroscopic datapoint, but also
calculates the $chi^2$ for the photometric and spectroscopic datapoints seperately.
This allows the user to weigh the spectroscopic data relative to the photometric data.
In practise the errorbars on the spectroscopic dataset are scaled (typically by a factor of order 0.2)
so that photometry and spectroscopy give roughly equal weight to the overall fit.
In the present model the dust temperature at the inner radius has been fixed to 1000~K, and we
assume a $r^{-2}$ density law, only fitting for the luminosity and dust optical depth (at 0.55~$\mu$m).
The outer radius is set to a few thousand times the inner radius, to where the 
dust temperature has reached 20~K, typical of the ISM.
MoD does not take into account the actual heating of the dust grains by 
the ISM and so this is an approximation.
Because of interaction of the expanding AGB wind with the ISM there can also 
be deviations for a $r^{-2}$ density law.
These approximations have no impact on the results as there is no far-IR 
data available for our sample that could constrain these values.
The longest wavelength data available for some stars is the PACS data at 
140~$\mu$m.
Some test calculations indicate the the flux in this filter is reduced by less 
that 10\% if the outer radius
were reduced by a factor of $\sim4$ below 1000 times the inner radius to where 
the dust temperature is about 30-35~K.

Several combinations of dust species have been tried to obtain a best fit.
They were olivine (amorphous MgFeSiO$_4$, optical constants from 
\citet{dorschner95}),
compact amorphous aluminum oxide \citep{begeman97},
and metallic iron \citep{Pollack1994}. The resulting abundance ratios for
each source are given in Table~\ref{tab:sed_model}.

Astronomical grains are not solid spheres and to mimic this the absorption and 
scattering
coefficients have been calculated assuming a "distribution of hollow spheres" 
(DHS, \citealt{Min03}) with a
maximum vacuum volume fraction of 0.7, that is appropriate for interstellar
silicate dust grains \citep{Min07_ISMgrains}.
An advantage of a DHS is that the absorption and scattering coefficients can be
calculated exactly for arbitrary grain sizes.
Single sized grains of 0.1, 0.2 and 0.5 $\mu$m have been considered.
The largest grain size used is inspired by recent observations of dust
around O-rich stars \citep{norris12, Scicluna2015, Ohnaka2016}.

The stellar photosphere was represented as a
MARCS model atmosphere\footnote{http://marcs.astro.uu.se/} \citep{Gustafsson2008}
of 2600~K (and $\log g$= 0.0, 2 \msol, and solar metallicity).
As shown below, all stars are so dust enshrouded that the SED fitting is insensitive to the input model atmosphere.

A canonical distance of 8 kpc has been assumed, slightly smaller than the value
quoted in the recent review by \citet{degrijsbono2016}, 8.3 $\pm$ 0.2 (statistical) $\pm$ 0.4
(systematic) kpc and which is based on an analysis of the up-to-date most complete
database of Galactic Centre distances.

The reddening law used in MoD is described in \citet{groenewegen12}.
The interstellar reddening $A_{\rm V}$ is taken from vH2007
%\cite{vanhollebeke2007}
(see Table~\ref{tab: sources}).

The model fits are shown in Figure~\ref{fig: seds} and the resulting parameters
(L$_*$, R$_{\rm in}$, $\dot{M_d}$, dust optical depth ${\tau}_{\rm V}$, grain size,
grain density $\rho$, and flux-weighted dust extinction coefficient $<Q>$)
in Table~\ref{tab:sed_model}.

Error bars are not listed explicitly as they are difficult to estimate.
The fitting returns the error on the parameters (luminosity and optical depth in this case).
These are typically very small as the resulting $\chi^2$ are large (reduced $\chi^2$ in the range 40-600).
This is related to the fact that the stars are variable and the SED is constructed by combining multi-epoch data,
without any attempt to average data in similar filters.
As the amplitude of the variability is (much) larger than the error on a single measurement this implies that the 
$\chi^2$ is typically always large.
One estimate for the error in luminosity and optical depth (hence dust MLR) comes from the internal error
scaled to give a reduced $\chi^2$ of unity.
A second estimate for the error comes from the values of the parameters in a range of $\chi^2$ above the best-fitting value.
This is required in any case, as for example
the absorption and scattering coefficients are external to the code, and the model is only run on a grid
with discrete values of the parameters (grain sizes and dust composition in this case).

Based on the above considerations, our best estimate for the 1~$\sigma$ error on the luminosity is 10\%, but that does not include the spread in distances
because of the depth of the Bulge ($\pm 1.4$~kpc, which gives a possible deviation of $\pm 35 \%$ in $L$).
The MLR scales linearly with the adopted distance.

As stated above, the errorbars on the spectroscopic datapoints is reduced by a certain factor as to give
all spectroscopic datapoints about equal weight in the fitting as all the photometry points.
Changing this scaling factor by a factor of two leads to a change of less than 10\% in the MLR and less than 1\% in luminosity.

The error on the optical depth is also of order 10\%, but the error on the dust MLR is larger.
This is related to the derived inner radius. The error on that quantity is 5\%, but there is a much larger error involved due
to the unknown effective temperature and dust temperature at the inner radius (both are hard to determine and have been fixed).
A realistic error on the inner radius would be a factor of 2, and this is then also a realistic error on the derived
dust MLRs.

The best fitting grain size is given (out of the considered values of 0.1, 0.2, and 0.5~$\mu$m), but none of the values
can be excluded. A larger grain size will lead to a higher dust extinction (less flux at shorter wavelengths), 
which could also be mimicked by a larger interstellar extinction. 
The values of the flux-weighted extinction coefficient scale with the adopted grain size.
To evaluate the impact we redid the SED modelling of IRAS 17251--2821, assuming 
a 0.1~$\mu$m grain (c.f. 0.5~$\mu$m grains in our best fit model). The 
luminosity remained unchanged, but the inner dust radius
decreases from 12.2 to 9.1 $R_*$ and the dust mass loss rate increases by 40\%.

\beginthetable
%\begin{table}
\centering
\caption{SED modelling parameters. }
\begin{tabular}{lccccccccc}
\hline 
  &  & & &  &  &  & & \\
IRAS & L$_*$  & R$_{\rm in}$  & $\dot{M}_{\rm d}$ & grain size & dust mix & $<Q>$ & $\rho$ & ${\tau}_{\rm V}$ \\
name & ($L_\odot$) & ($R_*$) & ($10^{-8}$ M$_{\odot}$ yr$^{-1}$) & ($\mu$m) &  & & (g cm$^{-3}$) & (at 0.55~$\mu$m) \\
\hline 
  &  & & &  &  &  &  & \\
17251-2821  & 4780 & 12.2 &  9.8 & 0.50 & MgFeSiO$_4$:Al$_2$O$_3$:Fe $=$ 90:10:10 & 0.25  & 2.70 &  64.5 \\
17276-2846  & 5120 & 13.6 & 20.6 & 0.50 & MgFeSiO$_4$:Fe $=$ 100:10 & 0.20  & 2.65 & 119.0 \\
17323-2424  & 4960 & 12.1 & 13.3 & 0.50 & MgFeSiO$_4$:Fe $=$ 100:30 & 0.32  & 3.45 &  90.8 \\
17347-2319  & 3880 &  9.2 &  6.6 & 0.20 & MgFeSiO$_4$:Al$_2$O$_3$:Fe $=$ 95:5:10 & 0.10  & 2.68 & 106.0 \\
17382-2830  & 5460 & 13.5 &  7.7 & 0.50 & MgFeSiO$_4$:Al$_2$O$_3$:Fe $=$ 90:10:3 & 0.23  & 2.35 &  41.7 \\
17413-3531  & 4920 &  9.0 &  2.5 & 0.50 & MgFeSiO$_4$:Al$_2$O$_3$:Fe $=$ 80:20:30 & 0.50  & 3.52 &  23.0 \\
17521-2938  & 4110 & 13.8 & 20.0 & 0.50 & MgFeSiO$_4$:Fe $=$ 100:10 & 0.20  & 2.65 & 127.0 \\
18042-2905  & 4600 &  7.9 &  4.3 & 0.10 & MgFeSiO$_4$:Fe $=$ 100:30 & 0.075 & 3.45 &  37.7 \\
\hline
\end{tabular}
\vfill
Notes: The dust MLR $\dot{M_d}$ is 
determined for a 10~\ks\ expansion velocity. $<Q>$ is the flux-weighted
extinction coefficient and $\rho$, the grain density. 
\label{tab:sed_model}
%\end{table}
\endthetable
\beginthetable
%\begin{table}
\centering
\caption{CO modelling parameters. }
\begin{tabular}{lccccccc}
\hline
 &  & &  & & &  \\
IRAS & $v_{e}$       & $v_{\rm drift}$  & $\dot{M_{\rm d}}$      & $r_{\rm gd}$ & $\dot{M}$            & $\epsilon$ \\
name & (km s$^{-1}$) & (km s$^{-1}$)  & ($10^{-7}$ M$_{\odot}$ yr$^{-1}$) &   
          & ($10^{-5}$ M$_{\odot}$ yr$^{-1}$) & \\
\hline 
 &  & &  & & &  \\
17251-2821  & 17.9 & 6.3 & 2.4 & 167 & 4.0 & 0.72 \\
17276-2846  & 17.2 & 4.4 & 4.5 & 195 & 8.7 & 0.75 \\
17323-2424  & 15.8 & 5.5 & 2.9 & 324 & 9.5 & 0.75 \\
17347-2319  &  9.0 & 2.8 & 7.8 & 385 & 3.0 & 0.75 \\
17382-2830  & 19.6 & 7.8 & 2.1 & 106 & 2.2 & 0.75 \\
17521-2938  & 18.2 & 4.5 & 4.6 & 154 & 7.0 & 0.75 \\
18042-2905  & 16.7 & 3.9 & 0.9 & 366 & 3.2 & 0.75 \\
\hline
\end{tabular}
\vfill
Notes: The dust MLR is corrected for the 
obtained dust velocity (see Section~\ref{sec:COmodel}). The $\epsilon$ 
parameter is the slope of the gas temperature power law, see Eq.~\ref{Eq-gas}.
\label{tab:co_model}
%\end{table}
\endthetable

\subsection{Modelling of the CO data}
\label{sec:COmodel}

To derive the gas MLR, we assume that the dust 
is driven by the radiation pressure and that the gas is driven by collisions
with the dust particles. For this, we solve the
equation of motion for dust-gas interaction based on \citet{Goldreich1976}. 
The dust MLR %which drives the wind 
and stellar parameters as well as the dust properties are taken
from the SED modelling in section \ref{sec:IRmodel}. We assumed that the 
stellar mass for these Bulge OH/IR stars is 2~M$_{\odot}$. The initial masses 
of the OH/IR stars will be further discussed in Section~\ref{sec:population}. 
In order to drive the gas to the observed gas terminal velocity, we input
the gas-to-dust mass ratio. Hence, we obtain the dynamical gas MLR for each 
object. At the same time, we calculate the dust drift velocity, $v_{\rm drift}$, 
via
\begin{equation}
 v_{\rm drift}^{2} = \frac{<Q> \, L_*\, v_{e}}{\dot{M}\, c}
\end{equation}
where $\dot{M}$ is the total MLR
%$Q$ is the dust absorption efficiency, 
%$L$ is the stellar luminosity, $\dot{M}$ is the MLR 
and
$v_{e}$ the gas velocity which is measured from the OH maser observations
(Table~\ref{tab: sources}) and corrected to the terminal velocity hence the gas 
velocity 
is 1.12$\times v_{\rm OH}$ (as described in \ref{sec: COdata}). The dust 
velocity is simply $v_{e} + v_{\rm drift}$. 
The dust mass loss required to fit the SED is then modified by the
derived dust velocity, $v_{\rm dust}$. The SED fitting measures the dust 
column density hence keeping $\dot{M}_{\rm dust}/v$ will maintain the 
overall SED fit. The modified dust MLR is again used as an input in the 
dynamical calculation in the iterative process to 
calculate the gas velocity and the dust velocity by modifying the dust-to-gas 
mass ratio.

The new value of the dust velocity is then used to calculate an updated dust
MLR (keeping the  $\dot{M}_{\rm dust}/v_{\rm dust}$ constant).

This 
iterative process is said to be converged when the values of the successive 
dust velocities agree to better than 1\%. Table~\ref{tab:co_model} lists the
parameters derived from the dynamical calculations.

In general, we can use the velocity profile to
probe the formation of lines with different excitation but in this study, 
CO J=2-1 and 3-2 arise in the region where the wind has reached its
final velocity and hence do not probe the acceleration zone.

We assume that the metallicity of the stars in the
Galactic Bulge is approximately solar \citep{Uttenthaler2015} and thus take a 
cosmic abundance of carbon and oxygen hence the CO abundance, CO/H$_{2}$, 
is 3$\times 10^{-4}$. We used the molecular radiative transfer code based on 
works by \citet{SchoenHemp} and \citet{Kay2004} to simulate the %observed APEX 
CO lines.
We took into account up to J=30 levels for both the ground and first
vibrational states of CO. The collisional rates for between the rotational
states in both v=0 and 1 are taken from \cite{Yang2010}.

We assume a gas temperature law in a form
\begin{equation}
 T_{\rm g}(r) = T_{\rm eff}/r^{\epsilon}
\label{Eq-gas}
\end{equation}
where $T_{\rm eff}$ is the effective temperature of the star and $\epsilon$
is a gas temperature exponent between 0.7 - 0.75
(see Table~\ref{tab:co_model})
which gives the best result for to the observed CO profiles.
We take into account the infrared pumping by the dust and assume a dust 
temperature in a form of

\begin{equation}
 T_{\rm d}(r) = T_{\rm con}/r^{\eta}
\label{Eq-dust}
\end{equation}
where $T_{\rm con}$ is the dust condensation temperature, 1000~K  and $\eta$ is
a dust temperature exponent of 0.45 - a slope derived from a single power
law from the dust SED modelling. We set the CO outer radius to be at %least 
1.5 times the CO photodissociation radius set to be where the CO
abundance drops to half its initial value \citep{Mamon1988}. Since the time this
analysis was performed, one of our co-authors presented a new paper on the
calculation of the CO photodissociation radius \citep{Groenewegen2017}, based on 
improved numerical method and updated H$_2$ and CO shielding functions. Taking the
resulting radius for a star with a MLR of $5 \times 10^{-5} M_{\odot}$ yr$^{-1}$ 
from his Table~1, as a representative case for our sample, we find a
radius which is 10\% smaller than what we used, on basis of \cite{Mamon1988},
well within the uncertainties.

The CO line intensities derived from our dynamical modelling are plotted
together with our CO data in Figure~\ref{Fig-CO}. It should be mentioned that at
these high MLR ($> 10^{-5} M_{\odot}$ yr$^{-1}$) the optically thick spectral 
lines become saturated \citep{Ramstedt2008}, so that the line intensities become 
less dependent to the MLR. This has no effect on our derived MLR as this is
obtained from the dynamical modelling.
The calculated CO lines are more sensitive to the gas kinetic temperature
  which is described by eqn.~\ref{Eq-gas}.

The dynamical MLR 
is derived from the assumption that the dust driven wind varies smoothly as 
1/$r^2$ for a spherical symmetric wind outside the acceleration zone where it 
has reached a constant terminal velocity.
Contrary to the modelling of OH/IR stars by \citet{just96},
there is no need to shorten the CO outer radius. For the sample
of our study no significant change in the MLR 
is required to fit the CO profile. We will further discuss this issue
in Section~\ref{sec: superwind}.

The derived gas-to-dust mass ratios range from about 100 to 400 and  
reflect a large range seen in galactic objects \citep{Kay2006}.
The derived MLRs are reasonably moderate for OH/IR stars and lower 
than those derived from galactic extreme OH/IR stars
which show MLRs in excess of 10$^{-4}$ M$_{\odot}$
yr$^{-1}$. The latter stars are thought to be intermediate-mass stars with 
initial masses M$_{\rm init} > 5$ M$_{\odot}$ based on their low $^{18}$O/$^{17}$O ratios 
\citep{Justtanont2015}.

In order to check how the input parameters affect the outcome, we changed the 
velocity by $\pm$15\% and calculate the resulting dust and gas mass loss rates. 
Changing the gas velocity by 15\% changes the dust mass loss rate and the dust 
(gas $+$ drift) velocity by the same amount but affects the gas-to-dust by 25\%. 
The combined changes result in a change in the derived gas mass loss rate by 
$\sim$ 10\%. 

Just like in Section~\ref{sec:IRmodel} we investigate here the effect of using 
a smaller 0.1~$\mu$m grain in our
modelling for IRAS 17251--2821. With the smaller grain size we find a smaller
$v_{\rm drift} = 2.2$~km s$^{-1}$ in the dynamical modelling and hence obtain a
difference of only 15\% in $\dot{M_{\rm d}}$, rather than the 40\% we obtained
in Section \ref{sec:IRmodel}. The total MLR becomes $5.2\, 10^{-5} 
{\rm M}_{\odot} {\rm yr}^{-1}$, i.e. 30\% higher than in the 0.5~$\mu$m grain case and 
the $r_{\rm gd}$ becomes 190 vs 167 (see Table~\ref{tab:co_model}).

\subsection{Periods of the variables}
\label{sec:Periods}
Amongst the Long Periodic Variable stars, OH/IR stars are known to have the largest
amplitudes ($\sim$ 1 magnitude bolometric) and the longest periods (several 
hundreds up to more than a thousand days).  vH2007
monitored the stars in near-infrared bands ($J,H,K$ and $L$). For four sources the 
period of variability could be established which are indicated in 
Table~\ref{tab: periods}. The other sources showed either variability, but no 
period could be established or were not detected in the K band. More recently, 
multi-epoch
observations from the VVV survey  in the K-band (we used public data from DR4) and the 
{\it AllWISE Multi epoch Photometry Table} and the
{\it Single Exposure (L1b) Source Table} from the NEOWISE reactivation mission
\citep{Mainzer2014} became available. For the latter we used data in the W2 filter at 
4.6~$\mu$m where the OH/IR stars stand out as bright stars with respect to the
surrounding stars. We did not use the longer wavelength filters W3 and W4, as  
the PSF becomes larger and increases the risk of crowding. We only used 
data with individual error bars less than 0.04 mag. 

The K- and W2-band data were investigated
separately to determine the periods, amplitudes and mean magnitudes using the 
program  {\it Period04} \citep{Lenz2005}.

The periods, averaged magnitude over the light
curve and the amplitude are shown in Table~\ref{tab: periods}.
For IRAS~17251--2821, which is very weak at K (14.5 mag), we were able to 
establish a period on the basis of the K-band VVV survey data where previously 
vH2007 could
only establish that the source was variable but could not determine a period. For the 
four sources with periods determined in vH2007
and now from 
the VVV-survey we find similar periods, only deviating by a few percent, except 
for IRAS~17347--2319 (see Section~\ref{sec:Indiv}). The vH2007 monitoring 
period took place in 2004 -- 2006 and the VVV data covers 2011 -- 2013, so that 
slight changes might be real.
The average $K$ magnitudes are generally fainter for the VVV survey than in the
vH2007 result, which can be explained by the difference in the
filter profile of the MSSO K band \citep{McGregor94} and the K$_{\rm s}$ band used in the
VISTA system \citep{Minniti2010}, combined with the very red SEDs of our sources. 
IRAS~17347--2319 and 18042--2905 show consistent K-band amplitudes, 17382--2830 and 
17413--3531 show much smaller amplitudes in our new fitting. For the latter source 
this might be related to the lower quality of the VVV photometry and 
subsequently of our fit.

For all our sources we were able to determine periods from the WISE survey. The
fact that our sources are brighter at 4.6~$\mu$m and suffer less of source
confusion is likely to explain this higher success rate. The periods are 
consistent with what is derived from the VVV K-band survey. 
In case of the WISE data the period from vH2007 or the period determined from analysing 
the VVV data was used as a first guess and the program was allowed to converge.
In the cases that there was only WISE data available, several periods were tried. 
For the further analysis, we adopt one period per source. In case of 2 K-band periods, 
we use the average value, for IRAS 17251--2821 we adopt the VVV derived period 
and in all other cases we take the WISE band derived value. 
The adopted periods are given in the last column of Table~\ref{tab: periods}.

\beginthetable
\centering
 \caption{ Variability parameters.} % } 
\begin{tabular}{  l c c c c c c c c c c}
  \hline
 & & & & & & & & & & \\
   & vH2007 &     &  & VVV &  & & WISE &    &   \\
  IRAS name   &  Period &  < K >  & $\Delta$ K & Period & < K > &
  $\Delta$ K & Period &  < W2 > & $\Delta$ W2 &  adopted \\
              &  (days)        &  (mag)    &  (mag)     &  (days)     &  (mag) &  (mag) &  (days)    &  (mag)        &  (mag)     &   P (days)   \\
  \hline 
 & & & & & & & & & & \\
17251-2821  & -- & 14.47 &  & 693 (7) & 16.07 (0.11) & 0.98 (0.02) & 681 (3) & 5.74 (0.03) & 1.10 (0.05) & 693 \\
17276-2846  & -- & &  &  & & & 488 (15) &  7.35 (0.02) & 0.72 (0.03) & 488 \\
17323-2424  & -- & & &  & &  & 552 (5) & 6.53 (0.05) & 0.94 (0.06) & 552 \\
17347-2319  & 355 & 11.09 & 0.23 & 290 (3) & 11.64 (0.02) & 0.26 (0.02) & 292 (1) & 4.99 (0.04) & 0.86 (0.05) & 323 \\
17382-2830  & 594 & 11.56 & 1.23 & 629 (9) & 12.2 (0.2) & 0.78 (0.03) & 625 (6) & 4.38 (0.07) & 1.09 (0.07) & 611  \\
17413-3531  & 624 & 10.09 & 1.10 &  664 (18) & 11.22 (0.05) & 0.68 (0.06) &
639 (9) & 5.38 (0.03) & 1.03 (0.04) & 644  \\
17521-2938  & --  &  &    &  &   &      & 562 (3) & 7.20 (0.01) & 0.51 (0.02) & 562 \\
18042-2905  & 594 & 10.87 & 0.78 & 574 (5) & 11.71 (0.04) & 0.80 (0.04) & 556 (3) & 5.28 (0.03) & 0.86 (0.04) & 584 \\
\hline
\end{tabular}
\label{tab: periods}
\\
Notes: Periods and semi-amplitudes taken from \cite{vanhollebeke2007}  and 
newly determined on basis of VVV and WISE survey data (see text). Uncertainties
are given between brackets, but are not available in vH2007.
The last column  gives the period that is further used in the analysis.
\endthetable

\subsection{Comments on individual sources}
\label{sec:Indiv}
\begin{itemize}
\item IRAS 17382-2830: Only source where we could only detect the CO (3-2) 
transition, and not the (2-1). The overall noise of the CO (2-1) measurement
is not different from the other measurements but this source is significantly 
closer to the galactic plane (latitude approximately 1 degree). The background
subtraction is more problematic because of the interference from the 
interstellar CO gas \citep{Sargent2013}. The stellar velocity taken from the OH 
is at $-$51.2~\ks (LSR) which overlaps with a region designated by
\cite{Dame2001} as the Nuclear Disk, which may cause the stronger fluctuations
in the baseline between -70 and +10 \ks. 
This would also explain the fact that whereas the other stars have stronger CO
(2-1) than (3-2) detection, we here detect only the 3-2 transition, which is
less hampered by the ISM. 
\item IRAS 17413-3531: No CO emission was detected. It is the bluest source in 
our sample with the 9.7~$\mu$m band still partially in emission and has the 
lowest $\dot{M}_{\rm d}$ (Table~\ref{tab:sed_model}). The CO emission may 
be too weak for a detection in our survey.
\item IRAS 17347-2319: Only a weak CO detection but the star has a very red 
SED with a high $\tau_V$ and a strong silicate absorption band at 9.7~$\mu$m, 
indicating a high MLR. The star has only a relatively short 
period (355 days in vH2007 and 290 days in our analysis). This star will be 
further discussed in Section~\ref{sec:devper}.
\item IRAS 17276-2846: This is one out of three sources in common with the
sample studied by JEE15. They find a double-peaked SED for this source, where
the 'blue' peak below 2~$\mu$m is believed to correspond to the stellar
photosphere and the red part to the mass loss during the AGB. The star would
have now ended the AGB phase and has become a
so-called proto-planetary nebula. We do not follow this interpretation. Both the
modelling of the SED and the CO line strengths point to a present high MLR. 
The IRS spectrum still shows a very strong 9.7~$\mu$m absorption band,
which would disappear rapidly after the mass loss has stopped \citep{Kay1992}. 
On the basis of the WISE data we also find that the star is
variable with a large amplitude ($\Delta$W2 $= 0.72$ mag) and so likely still 
on the AGB. 
We believe that the 'blue' counterpart is not associated with the OH/IR star, 
but a nearby confusing source and was thus not further considered for our analysis. 
\end{itemize}

\section{Comparison with JEE15 on L$_*$, \mdot and r$_{gd}$}
\label{sec:comparison}

As described in Sections \ref{sec:IRmodel} and \ref{sec:COmodel}, the modelling 
of the observed SED and the CO 
measurements is a two-step process where first the
infrared observations are fitted. The resulting dust MLR is
used as an input to derive the gas-to-dust ratio, and hence the gas MLR and CO
density leading to CO (2-1) and (3-2) transition line strengths.  
The MLRs ranging from 10$^{-5}$ to 10$^{-4}$ M$_{\odot}$ are 
typically what is expected from OH/IR stars and are not extremely high as MLRs in 
excess of 10$^{-4}$ M$_{\odot}$ yr$^{-1}$ have also been found \citep{Justtanont2015}. 
Comparison of MLRs with other studies needs to be done with care as different modelling methods
and assumptions can lead to different estimates of the mass loss and gas-to-dust
ratios. Also, in many studies the modelling is done on either only
observed SEDs or have only CO measurements available. Here we 
compare our results with those of JEE15 on a larger sample of Bulge OH/IR stars.

The JEE15 modelling is restricted to the SED fitting, making use of the OH maser
observations to have an estimate of the expansion velocity. They use the DUSTY 
radiative transfer code \citep{Dusty1999} to determine the luminosities and MLRs. 
We have three stars in common in our samples. 
For IRAS~17251--2821 they give a range of luminosity 3~100 - 7~200~L$_{\odot}$ 
where we find 4~780~L$_{\odot}$. The MLR ranges from 1.7 to 3.4~$10^{-5}$ \msolyr\ 
in JEE15 versus our slightly higher value of 4.8~$10^{-5}$ \msolyr. 
For IRAS~17322--2424 we find a larger difference in the MLR: 4.2~$10^{-5}$ vs. our 9.5~$10^{-5}$ \msolyr\ and
comparable luminosities: 4~200 and 4~960~L$_{\odot}$, respectively. JEE15 quote 
an uncertainty of a factor of 2 for the MLRs of the individual sources. 
For IRAS~17276--2846, the third source that is in common in our samples, they do not 
give model results, as they believe that the source has left the AGB (see Section \ref{sec:Indiv}).

To get a broader comparison with their results, we also compare their average
values for their larger sample with ours. As mentioned in Section~\ref{sec: sample}, 
JEE15 selected a sample which includes brighter
IRAS sources than we have. They divide their sample into low- and
high-luminosity groups, where their division lies at 7,000~L$_\odot$. As our
stars all belong to the first group, we will only compare with 
the average values for the MLRs of the so-called low-luminosity group. 
The thirteen 'low-luminosity' sources in JEE15 have an 
average MLR of 2.7~$10^{-5}$~\msolyr\ with a standard deviation 
of 1.6~$10^{-5}$~\msolyr\ versus our $(5.4 \pm 3.0)$ $10^{-5}$~\msolyr. 
JEE15 also used their DUSTY results to derive the gas-to-dust 
ratio, based on the SED fit and the expansion velocity, when known from the OH maser profile. 
This results in a value of r$_{gd} = 44 \pm 20$, which is considerably lower than
the average value that we find of $242 \pm 113$.

The differences in total MLRs and gas-to-dust ratios between JEE15
and ours may be explained by the different assumptions and inputs 
used in the modelling of the SED. JEE15 make use of optical constants 
for amorphous cold silicates from \citet{Ossenkopf1992} and 
the standard MRN \citet{Mathis1977} dust size distribution with
$n(a) \propto a^{-3.5}$, where $n$ is the number density and a is the size of
the grains. The grain sizes were limited to 0.005 $\leq a \leq 0.25~\mu$m. Our
assumptions are described in Section~\ref{sec:IRmodel}. We make use of a
combination of dust species and a single grain size, selecting the
best fitting one from 0.1, 0.2 and 0.5~$\mu$m respectively. Six
out of eight sources gave a best fit with a grain size of 0.5~$\mu$m, so larger
than what was used in the JEE15 modelling. 

To illustrate the impact of the assumed dust properties used, we redid the 
modelling of IRAS~17347--2319, 
using the silicates from \citet{KayXander1992} with a grain size of 
0.2~$\mu$m and a specific density of 3.3 g\,cm$^{-3}$. 
The resulting total $\dot{M} = 1.60$ $10^{-5}$~\msolyr and a 
r$_{gd} = 202$ are both a factor of approximately two lower than the values 
given in Table~\ref{tab:co_model}. JEE15 made a comparison of their gas-to-dust ratio
with the modelling by \cite{Kay2006} of OH/IR stars in the Galactic Disk. They
conclude that the Bulge OH/IR stars are on the low side in comparison to the
values given by \cite{Kay2006} which range from 50 to 180. Assuming that the
factor 2 difference in the gas-to-dust ratio for IRAS~17347--2319 that we find 
between our present modelling and the modelling using the input from 
\cite{Kay2006}, applies to our entire sample, we find a
similar range of r$_{gd}$ as what was found for the Disk stars. 

\section{The characteristics of the observed sample}
%\subsection{Periods, luminosities and mass loss rates}
\label{sec: characteristics}
In this Section, we will investigate what we can learn about important
parameters for the understanding of the AGB evolution, taking advantage of
having a group of stars at relatively well known distance and originating of
the same stellar population. In the next Section we will then discuss what we 
can learn about this population of stars in the Bulge.
 
\subsection{Luminosity and Period distribution}
\label{sec: lumdist}
The luminosities obtained from the SED modelling range from approximately 4~000
tot 5~500~L$_\odot$ and are on average 4729 $\pm$ 521 L$_\odot$, assuming that all
sources are at the distance of the Galactic Centre, taken at 8~kpc. 
 
        \begin{figure}
        \resizebox{\hsize}{!}{\includegraphics{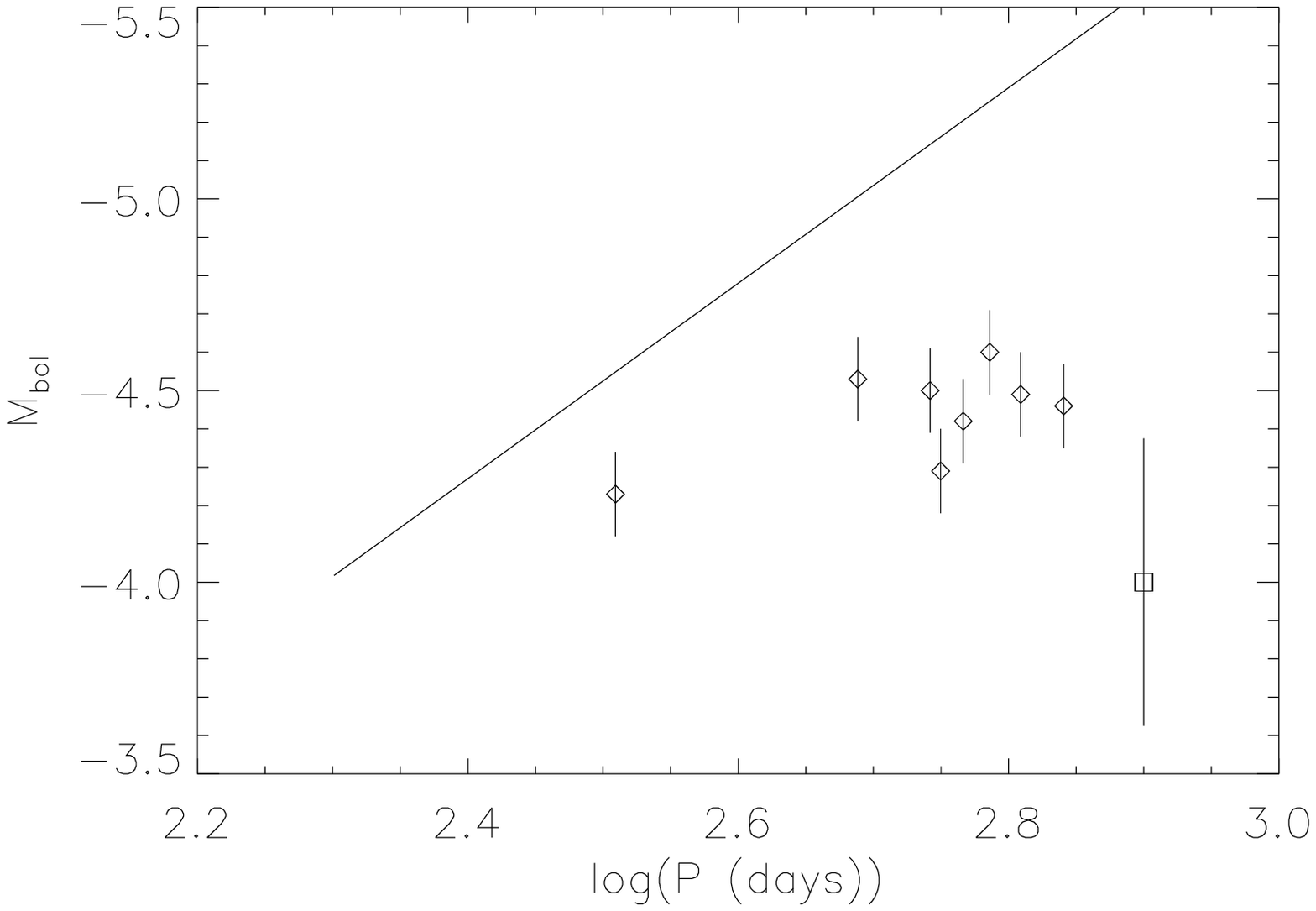}}
        \caption{  Period-Luminosity diagram for the OH/IR stars with
	the PL relation taken from \protect\cite{Whitelock1991}. The error bars 
	indicate the
	uncertainty from the SED modelling. The point at log P $=2.9$ gives the spread
	in luminosity because of the depth of the Bulge}
        \label{fig: PL}
        \end{figure}    
%	Whitelock et al 1991

The average luminosity of the sample agrees well with the peak in the 
luminosity distributions found by \cite{WilHarm1990} (5~000-5~500~L$_\odot$, for
a distance of 8.05~kpc) and by JEE15 ($\approx 4~500$~L$_\odot$, for a distance of 8~kpc). 
Here we want to point out that the well known and best studied OH/IR stars often have
luminosities well above 10~000~L$_\odot$ (e.g. \cite{beck10}), but that
\citet{Habing1988} in his analysis
of the galactic distribution of IRAS sources with OH/IR-like colours found a
luminosity distribution peaking at 5~000~L$_\odot$. The Bulge OH/IR stars are
thus not of exceptionally low luminosity.

The sample of OH/IR stars in JEE15 also contains stars with luminosities above
10~000~L$_\odot$ (their so-called high-luminosity group). We do not have these
because our selection of candidate OH/IR stars is based on the
\cite{WilHarm1990} sample as described in Section~\ref{sec: sample}. In the 
latter analysis of the flux distribution of IRAS
sources it is found that the stars with "apparant" luminosities above 
10~000~L$_\odot$ (F$_{12 \mu m} > 10$~Jy) are likely stars from the Galactic
Disk population. However, because of the lower number density, it could not be 
excluded that the Bulge also contained higher luminosity OH/IR stars. In our 
selection of stars to observe CO emission, we chose to select sources with the 
highest probability to be genuine Bulge stars and thus only selected stars with 
F$_{12 \mu m} < 10$~Jy. 

Most sources have periods in the range of 500 -- 700 days. This distribution
corresponds to the longest periods of the Bulge IRAS sources period distribution 
as determined by \cite{Whitelock1991}. As the OH/IR sources are the most extreme
AGB stars, this is no surprise. However, the periods are certainly not as extreme 
as several OH/IR stars in the Galactic Disk which have periods well 
above thousand days (e.g. \citet{vanlangevelde90}).

IRAS~17347--2319 has a clearly deviant period (P $= 323$~days) in comparison 
to the other OH/IR
stars and will be discussed in the Section~\ref{sec:devper}. 

\subsection{Period-Luminosity comparison}

Figure~\ref{fig: PL} shows the position of our stars in the so-called period-luminosity (PL) diagram. 
The full line shows the period-luminosity relation as derived by \cite{Whitelock1991} 
based on LMC oxygen-rich Mira variables with P$< 420$ days \citep{Feast1989} 
and Galactic Disk OH/IR stars with phase-lag distances \citep{vanlangevelde90}. 
This is the only PL relation that combines shorter period Miras with the 
longer period OH/IR stars that we are aware of. The distances 
towards the OH/IR stars in \cite{vanlangevelde90} were determined with the 
so-called phase-lag method 
and the overall uncertainty in $M_{\rm bol}$ is still between 0.5 and 1.0 mag. 
Clearly our OH/IR stars fall well below the relation as was also found for a 
sample of OH/IR stars in the Galactic Centre \citep{Blommaert1998} and 
we refer to the discussion in that paper on the PL relation for OH/IR stars. 
It should be noted that using an extrapolation of the PL relation of the oxygen-rich Miras in the LMC \citep{Feast1989} 
would only show an even larger deviation with our
OH/IR stars than with the PL relation used in this analysis.  
 
Rather than considering the OH/IR stars as an extension of the Miras 
towards higher masses, we believe that the OH/IR stars in this sample are to be seen as an
extension of the Miras towards a further evolved phase, as will be discussed in
the next session. The PL relation is also used to derive distances to the OH/IR
stars, for instance in \cite{beck10}. Although usage of the PL relation
is often the only way to get an estimate of the OH/IR star's luminosity, our 
result shows that this can lead to significant and systematic overestimation of the luminosity.

\subsection{Mass-loss rates versus Luminosity}

\begin{figure}
        \resizebox{\hsize}{!}{\includegraphics{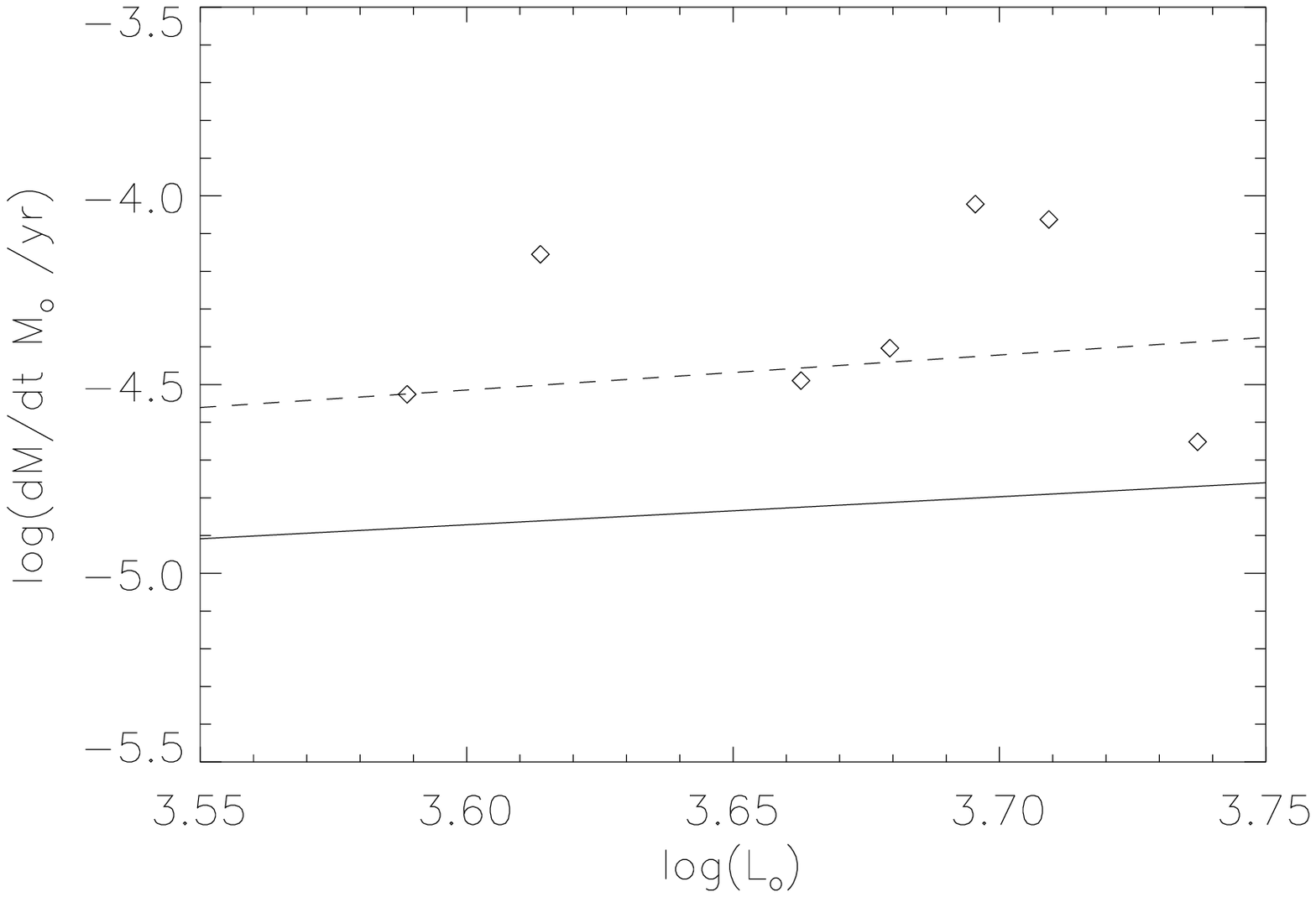}}
        \caption{   Mass-loss rate versus bolometric luminosity. The full
	line shows the classical limit of a single scattering event per photon. 
	The dashed line shows the empirical limit of MLRs that was suggested by
	\protect\cite{vanLoon1999} for a sample of oxygen- and carbon-rich stars in the
	LMC.}
        \label{fig: MdotvsL}
        \end{figure}

Figure~\ref{fig: MdotvsL} shows the MLRs of our stars versus the 
luminosities. In this
radiatively driven wind these quantities are not independent. The relation for
the so-called classical limit, i.e. only allowing one single scattering event 
per photon ($\dot M_{\rm classic} = L_* / (v_{\rm exp} . c)$), is shown in the 
figure.
\cite{vanLoon1999}  
showed that for a sample of AGB stars in the LMC several sources surpassed this
limit demonstrating that multiple scattering happens in dusty circumstellar
envelopes \citep{GailSed1986}. \cite{vanLoon1999} suggested a new empirical 
upper limit which is also indicated in Figure~\ref{fig: MdotvsL}.
Clearly all our sources surpass the classical limit, and three even surpass the
limit suggested by van Loon. 
The three sources with the highest MLRs also have the highest 
optical depths (Table~\ref{tab:sed_model}), where multiple scattering is likely 
to become increasingly important. Whether or not our stars indeed surpass the 
empirical limit suggested by \cite{vanLoon1999} is more difficult to answer, considering
the uncertainties one needs to take into account when comparing MLRs
derived from differerent methods (see Section~\ref{sec:comparison}). Lowering
the MLRs by a factor of two would bring the highest MLRs
just above the empirical relation given by \cite{vanLoon1999}. On the other
hand, the number of sources with such high MLRs in their paper is small and the
limit may be uncertain because of this. They also show that the optical depth 
by the
circumstellar shell is related to the $K-L$ colour. Our sources have redder $K-L$
colours than the oxygen-rich stars in their sample and thus may indeed be more 
extreme than the sample of LMC stars studied by \cite{vanLoon1999}.  This
difference in optical depth may be related to the likely higher metallicity of
the Bulge OH/IR stars in comparison to the LMC stars. And could indicate that
difference in MLR observed is real and related to the different populations.

%van Loon et al 1999

\subsection{Mass-loss rates versus Period}

A first condition to start an efficient dust-driven wind is the levitation of 
the gas, caused by large amplitude pulsations, to regions above the photosphere 
where grains can form. Earlier studies on the mass loss showed the dependency of
MLRs and the pulsation period of the AGB star (\cite{vassi93} and 
references therein) \cite{SchOlofsson2001} and \cite{beck10}). In agreement 
with \citet{vassi93}, \citet{beck10} find an
exponentially increasing MLR with period, until a maximum level is
reached where the MLR no longer increases. \citet{vassi93} see the 
leveling off occuring at a period of 500
days, whereas \citet{beck10} find that the MLR remains constant  
from approximately 850 days onward (at ${\rm
log}(\dot{M}) = -4.46$, with  $\dot{M}$ in units of \msolyr). We will come back to the comparison with the 
\citet{vassi93} result in the next section. Applying the relation provided by 
\citet{beck10} for periods shorter than 850 days:

\begin{equation}
{\rm log}(\dot{M}) = -7.37 + 3.42 \times 10^{-3} \times P
\label{Eq-MP}
\end{equation}

gives MLRs significantly lower than our values by a factor ranging
from 4 to 44 with a mean of 18 (we have excluded IRAS~17347-2319 because of its 
very short variability period, see Section~\ref{sec:devper}). On the other
hand, the scatter around the relation in \cite{beck10} is quite large  (up to a
factor ten below and above the relation), so that 
our MLRs are not entirely inconsistent with the MLRs 
obtained in their analysis. We conclude however that the relation given in 
eqn.~\ref{Eq-MP} for periods below 850 days, is not applicable to our stars, 
but that they have MLRs which agree with the 'plateau' value of
$\dot{M} \simeq 3.4 \times 10^{-5}$~\msolyr, the region associated with the 
{\it superwind} by \cite{vassi93}.

        \begin{figure}
        \resizebox{\hsize}{!}{\includegraphics{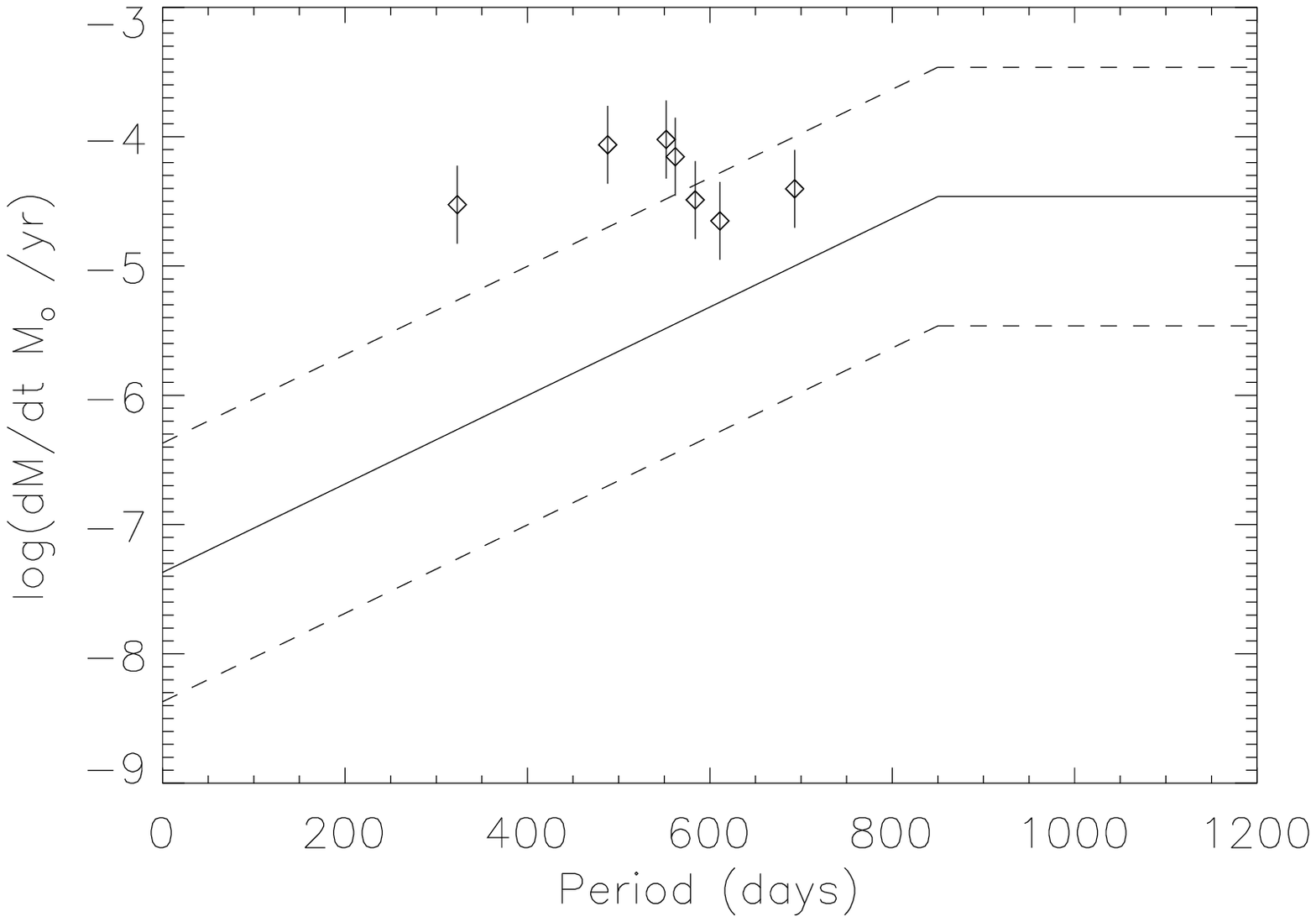}}
        \caption{  Mass-loss rate versus the variability period of our sources. 
	The full line shows the fit by \protect\cite{beck10} to the MLRs in 
	their sample for periods below 850 days. The MLR remains constant for
	longer periods. The dashed lines indicate a spread of a factor 10 around
	the fits, as is indicated in their Figure 14.}
	%The \dot{M} levels off to a 
	%constant value of $2.6 \times 10^{-5}$~\msolyr for longer periods.}
        \label{fig: MdotvsP}
        \end{figure}    
%De Beck et al. (2010)

\subsection{The deviant variability behaviour of IRAS~17347--2319}
\label{sec:devper}
IRAS 17347--2319 is standing out in the P-L diagram because of its short period 
in respect to the other OH/IR stars. The average period of
the other stars is 589 days, versus the adopted period of 323 we find for this
star. The full amplitude in the K-band is small (0.48 and 0.52 according to 
vH2007 and our own analysis, respectively) only just sufficient to be 
classified as a Mira variable (full $\Delta$~K $> 0.4$~mag, \citet{Feast1982}). 
Apart from the short period and small amplitude, this is also the only source 
which shows a deviating period between what is found by vH2007 and our own 
analysis (see Section~\ref{sec:Periods}). Where vH2007 finds a period of 355 days, 
we find consistently a period of 290 days from the VVV and WISE data. The data 
used in VH2007 was taken from  mid-July 2004 until November 2006, whereas for the VVV
and WISE survey we have data starting in April 2010 and ending in September 2013
for VVV and end of August 2015 for WISE respectively. A change in variability 
is noticed in
about one percent of the Mira variables \citep{Zijlstra2002,Templeton2005}.
\cite{Zijlstra2002} define three classes: "continuous change", "sudden change"
and "meandering change". In the first class, a continuous increase or decrease 
of the order of 15~\% occurs over a period of 100 years, whereas for the second,
such a change occurs ten times faster. In the third class, a
change of about 10~\% in the period duration is seen to happen over several
decades, followed by a return to the original period. The rapid change by 18\%
in less than a decade would place IRAS 17347--2319 in the second class, which
contains Mira variables like BH Cru, RU Vul and T Umi \citep{Uttenthaler2011}.
The occurance of a thermal pulse, when a helium-burning shell takes over from
the hydrogen-burning shell as the main energy source on the AGB, 
was suggested to explain the rapidly changing period \citep{Wood1981, 
Uttenthaler2011}. This could occur either in the build-up towards 
or in the aftermath of the thermal pulse. The strongly changing radius and
temperature of the star during such a thermal pulse will lead to a change in 
period ($P \propto R^{1.94} / M^{0.9}$), luminosity and expansion velocity 
\citep{vassi93}. An alternative explanation that has been suggested for a rapid 
decrease is a pulsation mode switch from the fundamental (assumed
to be the case for Mira's and OH/IR stars) to a low overtone mode (like in Semi
Regular Variables) \citep{LebWood2005}. 

Because of the large amplitude variability of this type of stars, it is not
possible to find evidence of a changing average luminosity of IRAS~17347--2319
over the time for which we have photometry available. However, the star has 
the lowest luminosity in our sample and is about 20\% lower than the average 
value of our sample. Also the expansion velocity significantly deviates from 
the average velocity found for the other stars: 9.0 vs 17.6 $\pm$ 1.3~km
s$^{-1}$. 
If  IRAS~17347--2319 is in a post-thermal pulse phase, one could also expect a
decrease in the MLR to occur \citep{vassi93} as is also observed in R
Hya, a Mira variable which has decreased its period from
500 to 385 days over a time period of about 300 years \citep{Zijlstra2002}. 
Such a MLR change is not clear from
our analysis. The SED shows a very strong obscuration in the visible and
near-infrared wavelengths, together with a strong silicate absorption band 
indicating a high present dust MLR. In the case of a rapidly
decreasing MLR, the stellar source would re-appear rapidly and the
silicate band would go into emission \citep{Kay1992}. 
We have much less information for
IRAS~17347--2319 than for BH Cru, RU Vul and T Umi, which have been monitored
for decades, to confirm that it is 
undergoing a 'sudden change' in variabilty. Further follow-up of the variability
and its SED would be highly desirable as this may be the first OH/IR type for
which such a behaviour has been observed.  

\section{the Bulge OH/IR population}
\label{sec:population}

Generally OH/IR stars are associated with stars of a few solar masses. Typical 
such examples are stars like OH26.5$+$0.6 which are very bright and have been 
studied in considerable detail \citep[e.g.,][]{Kay2006, Justtanont2015, 
groenewegen12}.
However, the OH/IR stars in our sample show much lower luminosities, typically in 
the range of 2000 - 7000 $L_\odot$ and periods below 700 days, whereas the more
luminous OH/IR stars reach periods well above a thousand days and luminosities
of several tens of thousands times the luminosity of the Sun. In contrast to
a number of extreme OH/IR stars, which are associated with active
star-forming regions in the galactic plane like the Molecular Ring, 
the OH/IR stars in the Bulge likely evolved from lower initial masses and are 
older, but still of intermediate age (1--3 Gyr). To
investigate this claim we will now make a comparison with the \cite{vassi93}
(hereafter VW93) evolutionary tracks.

 \begin{figure*}
\begin{minipage}{0.7\textwidth}
\resizebox{\hsize}{!}{\includegraphics{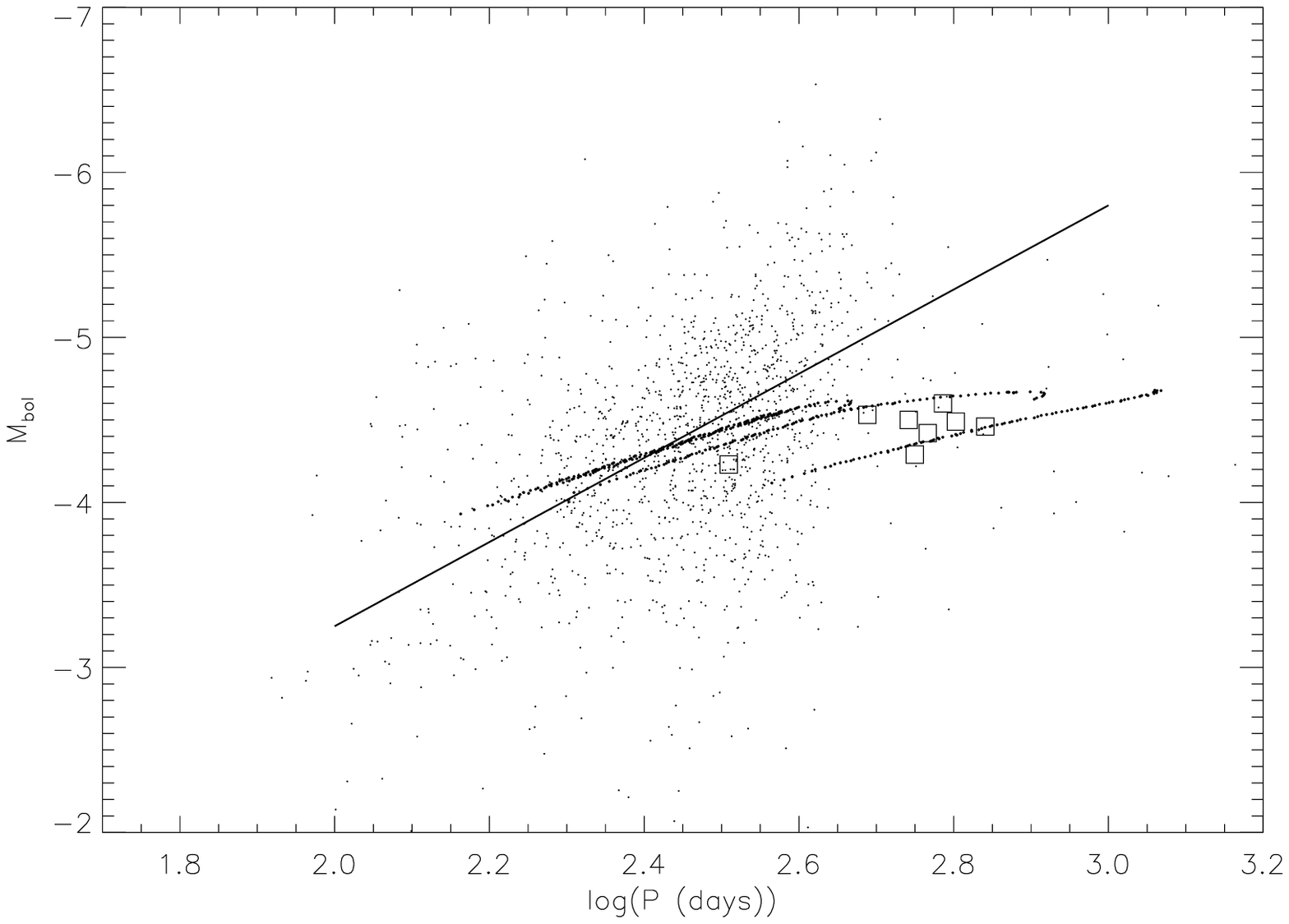}}
\end{minipage}
        \caption{ The PL diagram for Bulge Miras (taken from \protect\cite{GroenBlom2005}) 
	and OH/IR stars, together with the PL relation 
	(\protect\cite{Whitelock1991}; full line) and the evolutionary
	track from VW93 for a 1.5~M$_\odot$ star with a solar metallicity
	(H-burning phase of the last three thermal pulse cycles; dotted line).
	 }
        \label{fig: PL_VW}
        \end{figure*}    
%Groenewegen \& Blommaert 2005, Whitelock 1991

\subsection{Comparison with VW93}
\label{sec:VW93}

\cite{GroenBlom2005} studied the Galactic Bulge Mira variables on basis
of the OGLE-II survey and near-infrared photometry from the DENIS and 2MASS
all-sky databases. They show that the period distribution for stars within
latitudes ranging from -1.2$^\circ$ to -5.8$^\circ$ are indistinguishable and 
can be explained by a population with initial masses of 1.5 - 2~M$_\odot$, 
corresponding to ages of 1 to 3 Gyr. This result was based on synthetic AGB 
evolutionary models
where the synthetic AGB code of \cite{WagenGroen1998} was fine-tuned to
reproduce the models of VW93. VW93 provide calculations for a range of
metallicities, where \cite{GroenBlom2005} selected the Z = 0.016 model, for a 
solar mix. Studies of the metallicity of non-variable M giant stars in the Bulge
give a
slightly sub-solar value \citep{RichOriglia2005,Rich2007,Rich2012}. The same
metallicity was found for a sample of variable AGB stars in the Bulge by
\cite{Uttenthaler2015}. There are no direct measurements of the metallicities of
the OH/IR stars, but assuming that they originate from the same population as
the Bulge M giants and AGB stars, the solar mix is indeed the most appropriate.
The other values included in the VW93 models: Z = 0.008, 0.004 and 0.001, 
correspond to  the LMC, SMC and lower metallicity populations.

According to VW93 models, the maximum $M_{\rm bol}$ at the Thermal Pulsing AGB 
during H-burning are $-4.03$, $-4.52$, $-4.90$~mag for stars with solar metallicity and initial masses 
of 1.0, 1.5 and 2.0 M$_\odot$ respectively. The model for the 1.5 M$_\odot$ 
agrees closely to the Bulge OH/IR stars luminosity (Section~\ref{sec: lumdist}). In 
Figure~\ref{fig: PL_VW} we show the VW93 track (dashed line) for a star with 
M$_{init} = 1.5 M_\odot$ and
solar abundance in the PL diagram. The track covers the three last thermal pulse cycles before terminating
the AGB (with a duration $\approx 300,000$ years). 
We only include the part of track where the luminosity is produced by hydrogen 
burning, excluding the thermal pulses. Also shown are the 
Miras used in the \cite{GroenBlom2005} analysis and our OH/IR stars. The
large spread in the Miras can be explained by a larger spread in distance
(including fore- and background sources) and the single-epoch K band photometry
which was not corrected for variability. The bolometric magnitudes for the 
Mira stars were determined using
relation B of \cite{Kerschbaum2010}. The track follows the PL relation for a
large fraction where we also find the Mira stars. However, due to the changing 
mass, when the stars enter the so-called superwind phase, the period 
($P \propto R^{1.94} / M^{0.9}$) keeps increasing while the luminosity (related 
to the core mass) stays almost constant. The track then overlaps with the
position of the OH/IR stars in our sample.

Assuming that a star with $M_{init} = 1.5$ $M_\odot$ still needs to lose 
approximately 1~$M_\odot$ of material 
in its final phase before ending as a white dwarf of about 0.5~$M_\odot$ 
means that on average this OH/IR phase will last 20,000 years. 
VW93 model predicts 98,000 yrs duration of the superwind.
The longer duration predicted by VW93 is possibly connected to the single
scattering 'classical limit' they impose on the superwind MLR. As 
was discussed in Section~\ref{sec:comparison}, the MLRs we find 
surpass the classical limit by at least a factor of two.
VW93 predicts a ratio of the optical visible thermal pulsing AGB over the superwind phase of 
0.135. \cite{Blommaert1992} compared the numbers of IRAS sources with optical 
Miras in the Bulge. For the sources with OH/IR like colours the ratio is 0.02, 
significantly lower than predicted in VW93, but in agreement with the shorter
superwind duration of 20,000 years. 

In Figure~\ref{fig: PL_VW} it can also be seen that the VW tracks allow longer
periods (even above thousand days) than what we find in our sample. This is 
however, only true for a short
phase near the end of the AGB, which is even more reduced in time when allowing
higher MLRs, as we find is the case for the OH/IR stars. 
 
\subsubsection{Link between OH/IR stars and Miras in the Bulge} 
Further evidence for the connection between the Mira variables and the OH/IR 
stars is found in their distribution in the Bulge. \cite{Whitelock1992, 
GroenBlom2005} find that Mira stars follow a ''Bar'' structure with a viewing
angle of approximately 4$5^\circ$. Based on a dynamical modelling,
\cite{Sevensteretal1999} find that the OH/IR stars in the inner Galaxy are 
members of the Galactic Bar with a
viewing angle of 43$^\circ$ agreeing with the distribution of the Mira stars.

Recently, \cite{Harm2016} suggested that the galactic bar OH/IR stars are formed
in the Molecular Ring, an active star forming region at $\approx$ 4~kpc from the 
galactic centre (l = $\pm 25^\circ$) which connects to the end 
of the galactic bar \citep{Blommaert1994,Hammersley1994}. The stellar kinematics of the stars at the
tips of the bar are equal to those of the star-forming regions at these
locations, indicating that stars formed in the molecular ring can easily become
part of the galactic bar structure. Such a scenario agrees with the fact that 
we find stars of intermediate age and with the fact that gas-to-dust ratios of 
our OH/IR stars show the same range as was found by \cite{Kay2006} for a sample 
of OH/IR stars, situated predominately in the Molecular Ring.
The gas-to-dust ratio is
believed to be inversely related to the metallicity of the stars
\citep{Habing1994},
so that we conclude that the metallicities of the Bulge OH/IR stars are similar
to the selection of OH/IR stars in the Disk studied by \cite{Kay2006}.   

One final remark about the ages of the AGB stars in the Bulge. In the last
decades, there was often much debate on how the Bulge could contain AGB stars
like Miras and OH/IR stars of intermediate age in a galactic 
component which is believed to contain only an old stellar population 
\citep{Renzini1994, KuijkenRich2002, Zoccali2003}. This led to
suggestions that the Mira population was the result of merged binaries
(descendents of blue straglers, \cite{Renzini1990}). In the last decade
there has however been growing evidence that at least a (small) fraction of the
Bulge stars is of intermediate age as was shown by \cite{Gesicki2014} on basis 
of planetary nebula, by \cite{Bensby2013} for metal-rich dwarf stars.
The appearance of Miras and OH/IR stars in the Bulge is not so controversial in 
view of these recent results.

\subsubsection{Lack of carbon-rich stars in the Galactic Bulge}

Finally we return to the fact that JEE15 also have higher luminosity ($> 10~000 L_\odot$) OH/IR stars 
in their Bulge sample. As stated in Section~\ref{sec: sample} we have not selected such stars for our sample as
their 'true' Bulge membership is uncertain.  \cite{WilHarm1990} in their 
analysis of the luminosty distribution of IRAS stars with OH/IR colours assume
that all stars with IRAS $F_{12} > 10$~Jy are disk stars. They cannot exclude 
that OH/IR stars with luminosities above 10~000~$L_\odot$ exist, but that these
form at most 2\% of the population. JEE15's analysis of the high-luminosity 
group, comparing the luminosities with predictions from stellar evolution
models, shows that these stars have evolved from stars with $M_i \approx 2.0 - 6.0 M_\odot$. 
JEE15 indicate that the lack of carbon stars 
in the Bulge region (\citep{Blanco1989}) or very low number as indicated in the
recent paper by \cite{Matsunaga2017}, imposes problems with 
the stellar evolution  models.  Stars with initial masses above 4~$M_\odot$ can remain
oxygen-rich because of hot-bottom burning (HBB), but stars in the mass range between 
2 and 4~$M_\odot$ are expected to convert to C stars because of the third 
dredge-up when carbon is brought from the nuclear burning region up to the 
photosphere via convection (\cite{MarigoGirardi2007,KarakasLatt2014}). If the 
mass range of AGB stars in the Bulge is indeed limited to less
than 2~$M_\odot$, it would solve the problem of non-occurance of C stars in
the Bulge region. We repeat that \cite{GroenBlom2005} do not find
evidence of  stars with $M_i > 2 M_\odot$ in fields with galactic latitudes
above 1.2 degrees. The comparison field at $l=b=-0.05^\circ$ indicated the
presence of a younger population with $M \approx 2.5 - 3 M_\odot$ and ages
below 1 Gyr. This field is however much closer the Galactic Centre and in a 
region called the Nuclear Bulge, which is believed to be still active in star
formation \citep{Launhardt2002}.

\section{The duration of the superwind}
\label{sec: superwind}

Our combined SED and CO modelling does not impose any limit on the duration of
the superwind (Section~\ref{sec:COmodel}). The outer radius of the CO shell is 
taken at 1.5 times the radius of the CO photodissociation 
through interstellar UV radiation field. This is in contrast to what is found
for other OH/IR stars like OH~26.5$+$0.6 where the MLR derived from 
fitting the SED and solving the dynamical equation of the dust driven wind give 
a high MLR which overestimated the observed low-J CO lines by an 
order of magnitude. A way to reconcile the derived dynamical MLR 
and CO observations is that the current MLR (measured by the warm 
dust) is higher than in the past (as seen in J=2-1 CO line). High-J CO lines 
observed with {\it Herschel} are consistent with a sudden increase in MLR in the past 
couple of hundred years \citep{Justtanont2013}. This result, based on CO 
observations, is confirmed by an independent study of the fortserite dust 
69~$\mu$m band of which the shape and peak wavelength are 
very temperature sensitive \citep{koike03, suto06}. \cite{devries14} studied a sample of 
extreme OH/IR stars, including OH~26.5$+$0.6 and confirms the short duration 
of the superwind of less than a thousand years. As is stated in \cite{devries14}, 
such a short duration is problematic as the stars cannot lose sufficient mass, 
for instance in the case of a star like OH~26.5$+$0.6 this would be a couple 
of solar masses. The superwind would need to be followed by a phase of even
higher MLRs \citep{devries15}. An alternative scenario would be one
where several phases of a few hundred years occur in which the MLR increases 
to values above 10$^{-5}$~\msolyr. Such a timescale hints to a
connection to the thermal pulse, which is the only event on the AGB with such a
duration. The so-called 'detached shells' around carbon stars are believed to be
the result of interaction of a high and faster moving wind (10$^{-5}$~\msolyr) with a slower one with a  2 orders of magnitude lower MLR 
\citep{Olofsson2000, Schoier2005}.
Strangely, no oxygen-rich AGB stars are known with detached shells although it
is expected that thermal pulses would increase the MLR in a similar way.  

An alternative explanation for the above described CO lines' behaviour 
could be a higher impact of the interstellar UV radiation than is assumed 
in the radiative-transfer modelling (see Section~\ref{sec:COmodel}). 
The outer radius of the CO gas is determined by the photodissociation and is 
based on the work by \cite{Mamon1988}. 
If the interstellar radiation field is underestimated in the modelling, the 
outer radii of the gas %of different  CO line transitions 
will be smaller, increasingly so for lower J-transitions, as is observed in 
the case of OH26.5$+$0.6. This interpretation may also explain why we do not 
need to limit the CO outer radii for our Bulge OH/IR stars. 
Recent work by \citet{Groenewegen2017} demonstrate the effect of the ISRF and show that on average 
a factor of 15 increase in the ISRF will lead to a three times smaller photodissociation radius.
It can be expected that at high latitudes in the Galactic Bulge, where our
OH/IR stars are situated, the UV radiation field is much weaker than
compared to active star forming regions where higher mass stars like 
OH~26.5$+$0.6 are situated. Clearly this alternative interpretation does 
not explain the spatial distribution of the forsterite dust \citep{devries14}.

\section{Conclusions}

We have presented the succesful detection of the CO (3-2) and (2-1) transition
lines for a sample of OH/IR stars in the Bulge. On basis of our modelling of the
observed SED and CO lines, we find that the stars have an average 
luminosity of 4729 $\pm$ 521 L$_\odot$ and the average MLR is $(5.4 \pm 3.0)$ 
$10^{-5}$~\msolyr. 
Such MLR is well above the classical limit, with a single scattering event per
photon, for the luminosities in our sample. The 
variability periods of our OH/IR stars  are below 700 days and 
do not follow the Mira-OH/IR PL relation \citep{Whitelock1991}.  This result 
shows that usage of the PL relation for the OH/IR stars can lead to significant
errors in the luminosity determination. In comparison 
with the VW93 evolutionary tracks, we find that the stars have initial masses 
of approximately 1.5~M$_\odot$, which corresponds well with the findings of
\citet{GroenBlom2005} for the Bulge 
Mira variables, confirming the connection between the two groups of stars. If
more massive OH/IR stars are rare in the Bulge this may explain the scarcity of 
Bulge carbon stars. 
We find that the gas-to-dust ratio ranges between 100 and 400 and is similar to 
what is found for galactic disk OH/IR stars.
Contrary to findings of bright OH/IR stars in the Disk, our modelling does not 
impose a limit to the duration of the superwind below a thousand years. 
IRAS~17347--2319 has a short period of about 300 days which may be further
decreasing. Rapid changes in the variability behaviour have been observed for
Miras and may be connected to the occurance of a thermal pulse. It would be
the first time that such behaviour is observed in an OH/IR star.

\section*{Acknowledgements}
%\begin{acknowledgements} 
Based on observations with the Atacama
Pathfinder EXperiment (APEX) telescope (Programmes 088.F-9315(A) and 
090.F-9310(A)). APEX is a collaboration between the Max Planck Institute for 
Radio Astronomy, the European Southern Observatory, and the Onsala Space 
Observatory.
This research has made use of the VizieR catalogue access tool, CDS,
Strasbourg, France. The original description of the VizieR service was
published in A\&AS 143, 23. This publication makes use of data products from the 
Wide-field Infrared Survey Explorer, which is a joint project of the University 
of California, Los Angeles, and the Jet Propulsion Laboratory/California 
Institute of Technology, funded by the National Aeronautics and Space 
Administration. 
KJ acknowledges the support from the Swedish Nation Space Board. LD acknowledges
support from the ERC consolidator grant 646758 AEROSOL. We thank the referee
for a careful review of our manuscript which improved the quality of this paper.

\bibliographystyle{mnras}
   \bibliography{references}

\end{document}